\documentclass{article}

\usepackage{arxiv}
\usepackage{framed}

\usepackage[utf8]{inputenc} 
\usepackage[T1]{fontenc} 
\usepackage{hyperref} 
\usepackage{url} 
\usepackage{booktabs} 
\usepackage{amsfonts} 
\usepackage{nicefrac} 
\usepackage{microtype} 
\usepackage{threeparttable} 
\usepackage{lipsum} 
\usepackage{graphicx}
\usepackage[numbers,sort&compress]{natbib}
\usepackage{doi}
\usepackage{xcolor}
\usepackage{svg}
\usepackage{todonotes}
\usepackage{longtable}
\usepackage{array} 
\usepackage{graphicx}

\title{Large Language Model Agent for User-friendly Chemical Process Simulations}

\author{ \href{https://orcid.org/0000-0003-0815-3135}{\includegraphics[
	scale=0.06
]{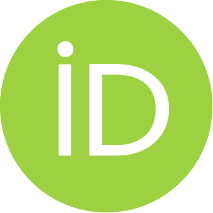}
\hspace{1mm}Jingkang~Liang}
\thanks{These authors contributed equally to this work.}
\\
Process and System Engineering Center \\
Department of Chemical and Biochemical Engineering\\
Technical University of Denmark\\
2800 Kgs. Lyngby, Denmark \\
\texttt{jinlia@kt.dtu.dk} \\
\And \href{https://orcid.org/0000-0003-0493-2270}{\includegraphics[scale=0.06]{
	orcid.pdf
}
\hspace{1mm}Niklas~Groll}\footnotemark[1] \\
Process and System Engineering Center \\
Department of Chemical and Biochemical Engineering\\
Technical University of Denmark\\
2800 Kgs. Lyngby, Denmark \\
\texttt{nigro@kt.dtu.dk} \\
\And \href{https://orcid.org/0000-0003-0513-4502}{\includegraphics[scale=0.06]{
	orcid.pdf
}
\hspace{1mm}G\"urkan~Sin} \\
Process and System Engineering Center \\
Department of Chemical and Biochemical Engineering\\
Technical University of Denmark\\
2800 Kgs. Lyngby, Denmark \\
\texttt{gsi@kt.dtu.dk} \\
}
\renewcommand{\shorttitle}{LLM Agent for Chemical Process Simulations}

\hypersetup{
	pdftitle={Large Language Model Agent for User-friendly Chemical Process Simulations},
	pdfsubject={ChemEng, LLM, Model Context Protocol},
	pdfauthor={Jingkang~Liang, Niklas~Groll, G\"urkan~Sin},
	pdfkeywords={First keyword, Second keyword, More},
}

\begin{document}
	\maketitle

	\begin{abstract}
        Modern process simulators enable detailed process design, simulation, and optimization; however, constructing and interpreting simulations is time-consuming and requires expert knowledge. This limits early exploration by inexperienced users. To address this, a large language model (LLM) agent is integrated with AVEVA Process Simulation (APS) via Model Context Protocol (MCP), allowing natural language interaction with rigorous process simulations. An MCP server toolset enables the LLM to communicate programmatically with APS using Python, allowing it to execute complex simulation tasks from plain-language instructions. Two water-methanol separation case studies assess the framework across different task complexities and interaction modes. The first shows the agent autonomously analyzing flowsheets, finding improvement opportunities, and iteratively optimizing, extracting data, and presenting results clearly. The framework benefits both educational purposes, by translating technical concepts and demonstrating workflows, and experienced practitioners by automating data extraction, speeding routine tasks, and supporting brainstorming. The second case study assesses autonomous flowsheet synthesis through both a step-by-step dialogue and a single prompt, demonstrating its potential for novices and experts alike. The step-by-step mode gives reliable, guided construction suitable for educational contexts; the single-prompt mode constructs fast baseline flowsheets for later refinement. While current limitations such as oversimplification, calculation errors, and technical hiccups mean expert oversight is still needed, the framework's capabilities in analysis, optimization, and guided construction suggest LLM-based agents can become valuable collaborators.
	\end{abstract}


	\section{Introduction}
\label{sec:introduction}


Chemical process simulators serve as an indispensable cornerstone of modern
chemical engineering practice, supporting engineers to design, analyze, and optimize
complex industrial systems. From petrochemical refineries to pharmaceutical
manufacturing plants \cite{suzuki2012simulation}, these sophisticated computational
tools enable engineers to predict process behavior, evaluate design alternatives,
and troubleshoot operational challenges before costly physical implementation \cite{aveva,aspen}.
However, despite decades of technological advancement, traditional simulation
workflows remain plagued by significant limitations that hinder their widespread
adoption and effectiveness.

The current state of process simulation is characterized by steep learning
curves, extensive manual configuration requirements, and the need for deep
domain expertise that can take years to develop
\cite{peter_review_2023,huang_deep_2025}. Engineers must navigate complex software
interfaces, manually select thermodynamic property models and operating
parameters, and possess intimate knowledge of underlying physical principles to achieve
meaningful results. This complexity creates substantial barriers, particularly for
junior engineers, interdisciplinary teams, and organizations with limited simulation
expertise, effectively restricting the accessibility of these powerful
mechanistic tools \cite{dahm_is_2002}.

Simultaneously, the artificial intelligence (AI) landscape has witnessed unprecedented
breakthroughs with the advent of Large Language Models (LLMs) and sophisticated AI
agent systems~\cite{ferrag_llm_2025}. These technologies have already
demonstrated remarkable success across diverse domains, from autonomous software
development and scientific research to complex problem-solving tasks that require
intricate tool integration and coordination \cite{chu_llm_2025,dong_survey_2025,yang_drugmcts_2025,xiao2024tradingagents}.
The emergence of agentic AI systems, capable of planning, reasoning, and executing
multi-step workflows, presents a compelling opportunity to address the
accessibility challenges of conventional chemical process simulators~\cite{rupprecht_multi-agent_2025}.

This combination of mature simulation technologies with cutting-edge AI
capabilities raises fundamental questions about the future of chemical
engineering practice: Can the sophisticated reasoning abilities of LLM-based agents
be effectively harnessed to assist (unexperienced) users in navigating complex simulation
workflows? How can we maintain the deterministic accuracy and safety standards
required in chemical engineering while leveraging the accessibility benefits of
conversational AI interfaces?

Addressing these questions, this work presents a novel framework that strategically
couple an LLM-based agent with a commercial chemical process simulation software
via a custom toolset. Rather than replacing traditional simulation methodologies
and workflows, our approach enhances them by providing an intelligent,
conversational interface that maintains the underlying rigor and reliability
that safety-critical applications demand. The proposed framework is designed to serve
as an assistant to experienced users in their daily tasks using the conventional
process simulator while simultaneously guiding new users through complex tasks of
analysing and synthesizing process simulation flowsheets, thereby broadening
access to advanced simulation capabilities without compromising engineering
standards.

The following sections of this paper are organized as follows: Section \ref{State
of the Art} reviews the current state of the art in LLMs, agentic AI systems, and
their applications in engineering domains, in particular, chemical engineering.
We highlight existing gaps and challenges and formulate this work’s aim. Section
\ref{sec:methodology} details the proposed agentic framework, including the
selection of an LLM agent, its integration with an agentic protocol, and its
integration with the commercial simulation software via a custom toolset. Section
\ref{sec:case_studys} presents two case studies testing the practical applications
of the framework in process analysis and synthesis, before Section
\ref{sec:results_and_discussion} discusses the results and implications of these
case studies. Finally, Section \ref{sec:conclusion} concludes with a summary of
key findings, implications for chemical engineering practice, and directions for
future research.

\section{State of the Art}
\label{State of the Art}


The development of LLMs has undergone a remarkable evolution, beginning with the
invention of the foundational transformer architectures by Vaswani et al.~\cite{vaswani_attention_2023}
in 2017 and culminating in sophisticated reasoning systems capable of complex
problem-solving \cite{wei_emergent_2022}. Early language models primarily focused
on text generation and completion problems, but the introduction of instruction
tuning~\cite{zhang_instruction_2025}, reinforcement learning from human feedback~\cite{bai_training_2022},
and chain-of-thought reasoning~\cite{wei_chain--thought_nodate} has dramatically
expanded their capabilities beyond simple text generation to sophisticated
reasoning and problem-solving tasks. The prevailing consensus among researchers and
practitioners suggests that we are witnessing the early stages of the LLM era \cite{matarazzo_survey_2025},
with ongoing improvements and breakthrough discoveries anticipated in the coming
years~\cite{zhang_evolving_2025}, particularly given the substantial computational
resources and capital investments being directed toward this field.

The latest step, featuring agentic AI systems, represents a fundamental paradigm
shift from monolithic language models to autonomous reasoning entities capable of
planning, executing, and reflecting on complex multi-step workflows, a critical
advancement for achieving universal applicability of LLM-based solutions
\cite{wang_survey_2024}. These AI agents leverage LLMs as cognitive cores while integrating
external tools, APIs, and environments to perform specialized tasks that extend far
beyond natural language processing~\cite{derouiche_agentic_2025}. Contemporary
agent architectures employ sophisticated planning algorithms, memory management systems,
and tool orchestration capabilities that enable them to decompose complex
objectives into executable subtasks, monitor progress, and adapt strategies
based on intermediate outcomes~\cite{yao_reac_2023,schick2023toolformer}.

To govern the operational integration of these autonomous entities, recently standardized
agentic protocols have been developed to define formal rules for system interoperability.
A primary example is the Model Context Protocol (MCP), launched by Anthropic in 2024~\cite{MCP}.
MCP provides a universal, vendor-neutral standard for connecting an agent's cognitive
core to external toolsets and data environments~\cite{MCP}. Unlike traditional
custom integrations, MCP enables a decoupled architecture where agents dynamically
discover and invoke capabilities through a unified interface. This
standardization supports both single-agent systems, which provide efficient,
centralized reasoning for direct task execution, and Multi-Agent Systems (MASs),
which distribute objectives across specialized entities to handle high-dimensional
problems through parallelized expertise. Single-agent architectures often remain
preferable for tasks that require lower latency and reduced computational overhead,
while MAS can offer greater robustness through peer-review loops \cite{du2024improving}.
Here, orchestration frameworks can play a crucial role in these configurations. Frameworks
such as LangGraph~\cite{langgraph} and AutoGen~\cite{autogen} provide infrastructure
for cyclic, stateful workflows, while Cognition~\cite{cognition} demonstrates specialized
autonomous engineering through integrated planning and environmental interaction.
Additionally, high-level protocols such as Agent2Agent~\cite{a2a_google_2024} can
help coordinate orchestration between these systems, providing an overarching
structure that ensures seamless integration and cooperation between diverse
agentic components. Together, these developments facilitate practical
applications of LLMs across technical domains, including tasks such as education,
software development, scientific research, and data analysis.

Frequently, LLM systems are developed first for educational purposes; LLM-driven
teaching assistants have been deployed to generate formative feedback, design learning
materials, and personalize instruction for students, ultimately enhancing learning
experiences and saving educators time~\cite{chu_llm_2025}. In software
development, agentic coding systems can autonomously plan tasks, write code, execute
tests, analyze failures, and iteratively debug programs until they succeed,
transforming developers' roles by reducing the mundane aspects of coding and allowing
them to focus on more innovative tasks~\cite{dong_survey_2025}. In scientific
research, LLM agents support multi-stage workflows by interpreting high-level research
goals, retrieving literature or data, designing experiments, and coordinating
external computational tools to implement partially automate research workflows, reducing researchers' workload and expediting discovery~\cite{ferrag_llm_2025}.
In drug discovery, MASs combine reasoning, retrieval, and search algorithms to
propose candidate molecules, evaluate properties, and refine designs across
iterative discovery loops, aiding scientists in accelerating the drug development
process and increasing efficiency~\cite{yang_drugmcts_2025,li_drugpilot_2025}.
Collectively, these applications demonstrate that LLM-agents are no longer
limited to text generation, but have evolved into autonomous, tool-using systems
capable of structured reasoning, planning, and multi-step execution across a
wide range of professional domains.

Also across the field of chemical engineering, efforts have been made to leverage
the power of LLMs and build agentic AI systems that support core activities.
Many integrations relate to chemical engineering education. Tsai et al.~\cite{tsai_exploring_2023}
highlights the potential benefits of integrating LLMs like Chat-GPT as a problem-solving
tool to enhance students' critical thinking; Caccavale et al.~\cite{caccavale_chatgmp_2025}
introduced ChatGMP, an LLM-based chatbot that supports student question answering by retrieving 
and synthesizing information from lecturer audio recordings using a retrieval-augmented generation framework.
Many other studies
similarly report that LLMs can support lab activities \cite{keith_harnessing_2025}, virtual tutoring \cite{caccavale_towards_2024}, and process-design
learning \cite{kong_complementary_2023}, reinforcing their growing value as
educational tools. Further, models
such as Coscientist~\cite{boiko_autonomous_2023}, ChemCrow~\cite{m_bran_augmenting_2024},
ChemLLM~\cite{zhang_chemllm_nodate}, and BatGPT-Chem~\cite{yang_batgpt-chem_2024}
enable autonomous experiment design, laboratory automation, retrosynthesis
planning, and multi-tool chemical reasoning, highlighting how LLM agents bridge
natural-language intent with computational chemistry workflows. In fault diagnosis,
systems like FaultExplainer~\cite{khan_faultexplainer_2025} leverage LLM
reasoning to interpret detected faults, explain sensor contributions, and
propose plausible root-cause hypotheses, thereby improving operator interpretability.
Finally, multiple studies explore LLM applications for chemical process design tasks,
such as process analysis and synthesis, including process simulations. Du and
Yang~\cite{r1} outline a next-generation agent architecture for intelligent
simulation workflows; Lee et al \cite{lee_gpt_2024} introduce a multi-agent GPT-based
system that generates process design improvements from structured flowsheet descriptions;
and Zeng et al~\cite{zeng_llm-guided_2025} demonstrate an LLM-guided multi-agent
optimization system that autonomously infers operating constraints and
outperforms traditional methods on hydrodealkylation optimization. In simulation
automation, Liu et al.~\cite{liu_toward_2025} show that autonomous simulation agents
can plan, execute, and analyze full research simulations, achieving multi-cycle
long-task execution with minimal human input, while Rajeev et al.~\cite{rajeev_application_2025}
integrate an AI assistant directly with the commercial platform AVEVA Process
Simulation (APS) to support engineers in developing and troubleshooting
industrial flowsheets. Collectively, these advancements indicate that LLM-based
technologies have the potential to transform multiple facets of the chemical
engineering lifecycle, including classroom instruction, research, industrial-scale
simulation, and process design.

Despite recent advancements, key challenges remain in developing trustworthy LLM-based
assistants that can truly support experienced chemical engineers in complex,
real-world tasks. Specifically, existing LLM-based technologies have not yet
addressed tasks requiring deep system understanding and decomposition into
interconnected subproblems, such as the analysis and synthesis of industrial production
systems \cite{seider_product_2016}. Chemical engineers rely on specialized
simulation software for rigorous solutions, but these tools often pose usability
challenges due to manual configuration requirements and the need for substantial
domain expertise. Thus, a fundamental gap exists in transforming traditional,
human-centric simulation software into flexible, conversational tools powered by
LLMs, while preserving the rigor and reliability necessary for safety-critical
engineering applications.

To address this gap, our work aims to develop and evaluate an LLM agent framework,
built on MCP, that enables natural-language interactions with the commercial simulator
APS. This framework is designed to ensure safe, reliable, and reproducible
execution by integrating a curated toolset that bridges conversational AI with
deterministic process simulation. We demonstrate the framework’s practical utility
through two comprehensive case studies, showcasing its application in process
understanding, analysis, optimization, and flowsheet synthesis. Finally, we critically
assess the strengths and limitations of LLM-driven agents in supporting chemical
engineering workflows, highlighting how such systems can make commercial process
simulators more accessible and efficient without sacrificing the rigor required
for industrial use.

\section{Methodology}
\label{sec:methodology}

\begin{figure}[htbp]
    \centering
    \includegraphics[width=0.4\textwidth]{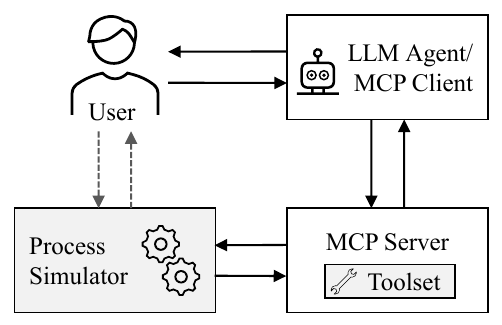}
    \caption{Schematic overview of the proposed framework including MCP.}
    \label{fig:overall_framework_screenshot}
\end{figure}

Our proposed framework, as illustrated in Figure
\ref{fig:overall_framework_screenshot}, intends to replace the direct, manual
interaction between a user and a commercial process simulator with an
intelligent, conversational interface powered by an LLM. Instead of navigating
the simulator's graphical user interface or writing custom scripts, the user communicates
with an MCP-enabled LLM agent, also called the MCP client, in natural language.
The LLM agent interprets the request, plans the necessary actions, and invokes a
curated toolset exposed by a dedicated MCP server. The server, in turn, interfaces
with the process simulator and executes all deterministic engineering operations.
This section details the framework's architecture, covering the MCP client, the
MCP server, and the integration with the process simulator, including the
curated toolset.

\subsection{LLM Agent and MCP Client}

The LLM agent, running within a host application, serves as the primary
interface to the simulation environment. It interprets natural-language requests,
decomposes complex tasks, orchestrates tool calls, and maintains conversational
context. This contextual reasoning is essential for ensuring that the system correctly
interprets user intent and provides accurate, relevant responses. Additionally,
the LLM agent should detect and handle user errors and present results in clear,
user-friendly natural-language outputs.

Selecting a suitable host application, LLM agent, and underlying model
involves careful consideration of several criteria, including strong reasoning
capabilities for multi-step tool planning, reliable tool-use behavior, and a
sufficiently large context window to preserve conversation history. For this work,
we chose Claude Desktop, powered by Claude Sonnet 4.0 (Anthropic), as our LLM
agent \cite{claude4}. This decision was motivated by its state-of-the-art performance
on reasoning tasks and tool-use benchmarks in technical and scientific domains \cite{lmarena}.

Within this architecture, Claude Desktop acts as the host application, instantiating
an MCP client responsible for protocol-level communication. The LLM interacts
with this client, instructing it to discover, connect to, and invoke tools exposed
by MCP servers. Because Claude Desktop includes native MCP client functionality,
this setup enables seamless integration with custom MCP servers without requiring
additional middleware or custom communication layers.

\subsection{MCP Server}

The MCP server acts as the intermediary layer between the LLM agent and the
process simulation software. By abstracting the simulator's Python API into a clean,
high-level toolset with clearly defined inputs, outputs, and usage descriptions,
the MCP server enables the LLM to interact with the simulator without requiring model-specific
customization. Because the LLM only accesses these abstract tool definitions, rather
than the simulator's internal API, the same toolset can be reused across
different LLMs or host applications, enabling seamless migration and future extension
of the framework.

Our implementation uses FastMCP, a modern Python framework for building
lightweight MCP servers~\cite{jeremiah_lowin_fastmcp_nodate}. FastMCP provides a
streamlined interface for defining tools, specifying request and response
schemas, and handling communication via standardized JSON-RPC messaging. Each
tool is implemented as a typed Python function annotated with metadata
describing its parameters, return types, and purpose. This design pattern simplifies
server development while providing automatic schema generation and robust tool discovery
for the host application.

The MCP server runs as a standalone process, and the host application (Claude Desktop)
connects to it through its built-in MCP client. Once connected, the LLM agent
can query available tools, invoke them with appropriate arguments, and receive
structured responses containing simulation data or execution results. This architecture
ensures a reliable, transparent communication pipeline, enabling the LLM agent to
coordinate simulation steps while deterministic computations are handled by the server
and the underlying simulator.

\subsection{Process Simulator Integration}

The connection between the MCP server and the employed process simulator, APS (version
2025), is established via the APS scripting interface, a comprehensive Python-based
API that provides programmatic access to all core simulation functionalities,
including flowsheet construction, parameter manipulation, and result extraction.
Our custom MCP server implementation wraps selected APS scripting interface functions
into a curated set of high-level tools. Each MCP tool encapsulates one or more
APS API calls, which are the building blocks of the tools, into a single
meaningful operation, aligned with common simulation steps a process engineer would
take.

The full toolset exposed by the MCP server is summarized in Table~\ref{tab:mcp_tools}.
It enables the MCP client to perform a wide range of simulation tasks, from simple
data queries and flowsheet inspections to the construction of new flowsheets by adding
models and connectors. The modular design of the MCP server architecture ensures
the toolset is easily extensible when additional functionality is needed, enabling
future expansion with capabilities such as automated sensitivity analysis,
optimization routines, or advanced convergence troubleshooting without requiring
changes to the underlying LLM agent configuration. Critically, the MCP server serves
as a semantic bridge, orchestrating high-level simulation logic through natural
language reasoning and tool selection. It translates conversational intent into rigorous
engineering actions, maintaining complete control over the deterministic
execution logic. At the same time, all actual computations remain firmly
grounded in the deterministic, physics-based solver algorithms provided by APS.

\begin{longtable}
    {>{\raggedright\arraybackslash}p{0.15\textwidth}>{\raggedright\arraybackslash}p{0.33\textwidth}>{\raggedright\arraybackslash}p{0.17\textwidth}>{\raggedright\arraybackslash}p{0.25\textwidth}}
    \caption{MCP server toolset for APS integration.}
    \label{tab:mcp_tools} \\
    \hline \textbf{Name} & \textbf{Description} & \textbf{Inputs} & \textbf{Outputs}
    \\
    \hline \endfirsthead

    \multicolumn{4}{l}{{\textit{Continued from previous page}}} \\
    \hline \endhead

    \hline \multicolumn{4}{r}{{\textit{Continued on next page}}} \\
    \endfoot

    \hline \endlastfoot

    aps\_connect & Connect to APS. This establishes a connection to the AVEVA
    Process Simulation system and initializes all available managers (simulation,
    model, connector, etc.). & - & connected: bool, [...] \\

    sim\_open & Opens an existing APS simulation. & sim\_name: str & success: bool,
    [...] \\

    sim\_create & Create a new simulation in APS. & sim\_name: str, [...] &
    success: bool, [...] \\

    sim\_save & Save an APS simulation. & - & success: bool, [...] \\

    sim\_status & Get input, specification, and convergence status of APS simulation.
    & - & success: bool, solved: bool, has\_required\_data: bool, properly\_specified:
    bool, [...] \\
    \hline

    models\_list & Get detailed information about all models present on the specified
    APS simulation flowsheet. & - & success: bool, count: int, models: List[Dict[str,
    str]], [...] \\

    connectors\_list & Get detailed information about all connectors present on
    the specified APS simulation flowsheet. & - & success: bool, count: int,
    \newline
    connectors: List[Dict[str, str]]
    \newline
    [...]\\

    model\_all\_vars & Get detailed variable information of all variables of one
    specific model, including variable type, current value, units of measurement,
    and description. & - & success: bool, variable\_values: Dict[str, str],
    \newline
    total\_count: int, [...]\\

    model\_all\_params & Get detailed variable information of all parameters of
    one specific model, including parameter type, current value, units of measurement,
    and description. & - & success: bool,
    \newline
    variable\_values: Dict[str, str],
    \newline
    total\_count: int, [...] \\
    \hline

    var\_get\_multiple & Get values of multiple simulation variables at once. & variable\_paths:
    List[str] & success: bool,
    \newline
    variables: Dict[str, Any], set\_variables: int, [...]\\
    \hline

    var\_set\_multiple & Set values of multiple simulation variables at once. &
    variable\_paths: List[str] & success: bool,
    \newline
    variables: Dict[str, Any],
    \newline
    set\_variables: int, [...]\\

    param\_set\_multiple & Set values of multiple simulation parameters at once.
    & parameter\_paths: List[str] & success: bool,
    \newline
    updated\_parameters: int,
    \newline
    results: Dict[str, Any], [...]\\
    \hline

    model\_add & Add a model to the APS simulation. & model\_type: str
    \newline
    x: float, y: float & success: bool, [...]\\

    models\_connect & Creates a connection between two model ports and validates
    that the connection is real. & start\_port: str, end\_port: str & success: bool,
    \newline
    connector\_name: str, [...]\\
    \hline

    fluid\_create & Create a new fluid model in APS with the desired components
    and thermodynamic settings. & library\_name: str,
    \newline
    fluid\_name: str,
    \newline
    components: List[str], thermo\_method: str,
    \newline
    phases: str & success: bool, [...]\\

    fluid\_to\_source & Set fluid type of one specific source model. & fluid\_name:
    str, source\_name: str & success: bool, [...]\\
\end{longtable}


\section{Case Study Applications}
\label{sec:case_studys}

To demonstrate the practical application of our proposed agentic framework, we select
an existing simulation model from the APS examples library as a baseline and comparison
flowsheet. The simulation example "C1 - Water Methanol Separation" models the
binary separation of methanol and water via distillation, and is detailed in the
next section. Based on this example, we design two case studies to test our proposed
workflow. The case studies target common tasks of process simulation engineers, the
(1) analysis of simulation flowsheets and (2) synthesis of simulation flowsheets
detailed in Sections \ref{sec:case_study_1} and \ref{sec:case_study_2}.

\subsection{Water Methanol Separation Process}

Methanol serves as a key feedstock for the production of various chemicals and
fuels making its production and purification critical processes in the chemical
industry \cite{bozzano_efficient_2016}. In particular, the mixture of water and
methanol is commonly encountered in industrial processes, for example in biodiesel
production, where methanol is used as a reactant and must be first recovered
from the aqueous phase to facilitate the downstream processing steps
\cite{burmana_improving_2025}.

For analyzing the process configuration, an understanding of the fundamental chemical
and physical principles is necessary. Although the boiling points of methanol (64.7
°C) and water (100 °C) at ambient pressure differ by approximately 35 °C, suggesting
a relatively straightforward separation, this binary system features some
interesting challenges. The water-methanol mixture exhibits non-ideal vapour-liquid
behaviour, characterized by positive deviations from Raoult's law, requiring accurate
thermodynamic models, such as NRTL, to capture phase equilibrium correctly \cite{kurihara_isobaric_1993}.
While the system does not form an azeotrope, the relative volatility varies significantly
with composition, affecting separation efficiency across the column height. Due to
these non-ideal behaviors, careful consideration of design (e.g., number of stages)
and operational (e.g., reflux ratio) parameters is required to achieve desired
product purities while maintaining energy efficiency. These characteristics position
the binary water-methanol separation problem as an ideal benchmark system for process
analysis and optimization. The system is simple enough for rapid analysis by
domain experts, yet sophisticated enough to test the LLM agent's understanding of
thermodynamic principles, equipment configuration, energy-efficiency considerations,
and process-optimization strategies in a realistic industrial context.

\begin{figure}[htbp]
    \centering
    \includegraphics[width=0.6\textwidth]{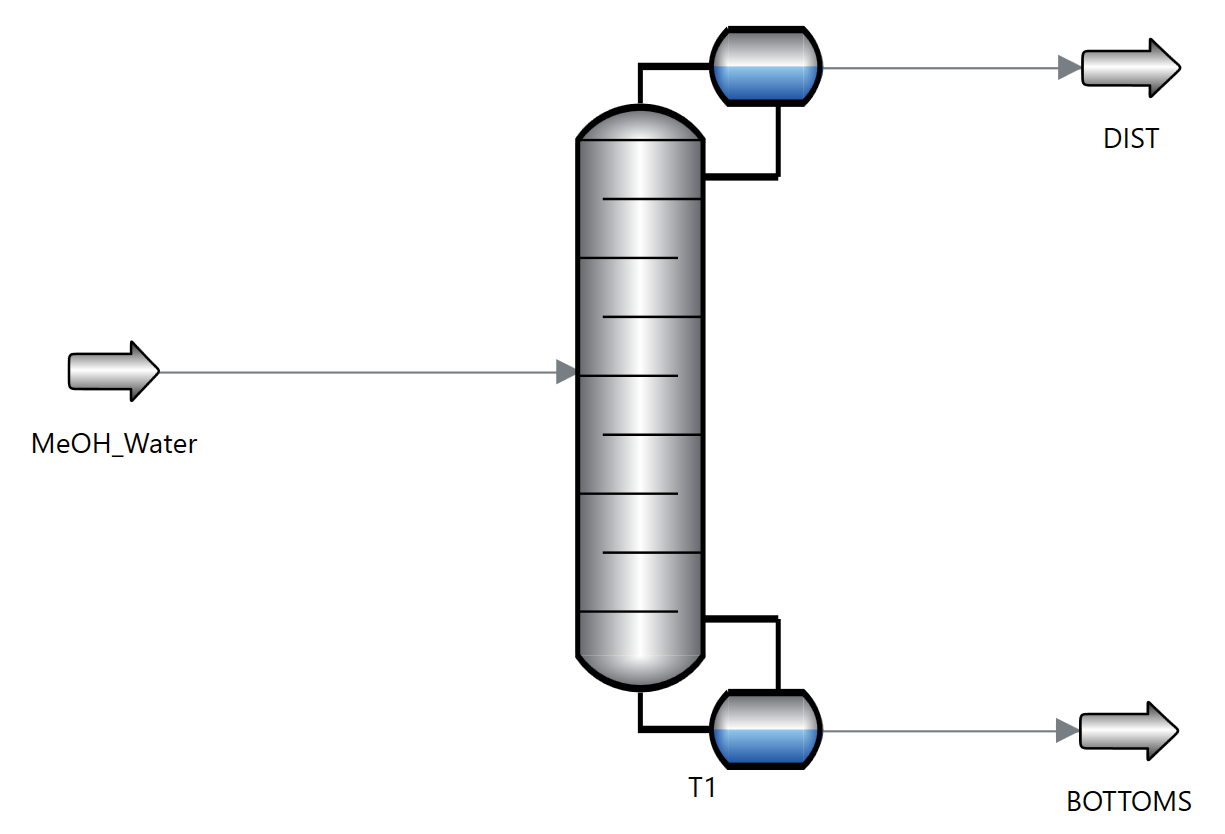}
    \caption{Simulation flowsheet of APS example "C1 - Water Methanol Separation".}
    \label{fig:flowsheet}
\end{figure}

The process simulation flowsheet for the water-methanol separation, shown in
Figure~\ref{fig:flowsheet}, is rather simple, consisting of a source stream, a distillation
column equipped with both a reboiler and a condenser, and sink modules for
collecting the bottom and distillate products. We define the feed stream as an equimolar
mixture of water and methanol, with a flow rate of 10 kg/s, at the specified operating
conditions (e.g., 80°C and 1 atm). Critical parameters that can be adjusted to
optimize the separation include reflux and reboiler ratios, the number of stages
in the distillation column, the feed stage location, and overall operating pressure
and temperature profiles. While the synthesis of the process requires careful
selection of appropriate modules, considering the underlying chemical and physical
principles to obtain a converged simulation flowsheet, the number of steps
required is limited, implying a fast flowsheet construction for experienced users.
Therefore, the binary water-methanol system provides a realistic yet manageable example
that captures the key challenges of synthesizing industrial separation processes.

\subsection{Case Study 1: Analysis of the Simulation Flowsheet}
\label{sec:case_study_1}

In the first case study, we aim to evaluate the LLM agent's ability to analyze,
interpret, and improve the existing water-methanol separation simulation modeled
in APS. The task is structured into two distinct steps. First, we ask for a
detailed summary of the modeled process, and second, we request a list of
potential process improvements. This case study should assess the system's capacity
to comprehend complex process simulation models, communicate findings
effectively to the user, and offer expert-aligned recommendations for system improvements.

The initial user prompt asks for a detailed summary of the specified simulation file.
To fulfill this request, we expect the agent to connect to APS, open the
relevant simulation file, and gather data on all present modules and their connections.
Further, it should independently identify and extract the values of relevant key
process variables and parameters. After screening all extracted information, we expect
to obtain a structured, detailed natural-language description of the simulated
process. The summary should include the flowsheet structure and key process specifications,
such as feed composition, operating conditions, and energy demands.

In the second step, the user asks the agent how to improve the simulated process.
Based on the previously extracted information, the agent can decide to extract more
information from the simulation model if needed. After that, the agent should provide
a summary of multiple points that could improve the separation task based on different
metrics, such as separation efficiency, purity of the product streams, or energy
demand. Dependencies between different variables and operating conditions should
be clearly stated. Finally, a comprehensive list of suggested improvements should
be provided to the user.

\subsection{Case Study 2: Synthesis of the Simulation Flowsheet}
\label{sec:case_study_2}

The second case study assesses the agent's process synthesis capability with the
primary goal of constructing an entire simulation flowsheet in APS from
high‑level user specifications. We evaluate and compare two construction modes based
on different user interactions and prompting: first, we test the assembly with a
stepwise dialogue between the user and the LLM agent; second, we provide the agent
with a single user prompt.

In the first approach, the user has almost full control, guiding the agent to assemble
the simulation flowsheet by incrementally requesting the addition of unit operations,
connection of units, and setting of parameter values. This approach mirrors
manual construction, but instead of manually clicking through the APS graphical user
interface, the user now provides the LLM agent with the required next step. This
way, the user can control that each step is conducted correctly and provide help,
or even take over a step, when the agent encounters problems. The user reduces the
complexity of each step by breaking the task into steps so that the agent requires
only a single tool call. Concurrently, the complexity can also be enhanced by combining
subtasks into the next user prompt. In this mode, we test the agent's ability to
follow stepwise user instructions to construct the simulation flowsheet in APS using
the provided toolset.

In contrast, with the second mode, we evaluate the agent's capacity to autonomously
construct a simulation flowsheet from a single user prompt. This mode requires
the agent to possess a sophisticated, abstract understanding of the problem statement
and to independently decompose the task into executable subtasks using the available
toolset. The process requires orchestrating multiple sequential tool calls with
progressively increasing information availability to fulfill this complex
objective. With this mode, we test the agent's ability to independently plan and
execute a multi-step workflow to synthesize a complete process simulation
flowsheet based on concise high-level user specifications.





\section{Results and Discussion}
\label{sec:results_and_discussion}

In the following, we present and discuss the results obtained from the two case
studies detailed before. The performance of the LLM-based agent in analyzing and
synthesizing simulation flowsheets in APS is qualitatively evaluated, including
the criteria correctness, completeness, efficiency, and user satisfaction. While
we show only excerpts in this section, all full prompts and answers can be found
in the appendix~\ref{app:prompts}.

\subsection{Case Study 1: Analysis of the Simulation Flowsheet}

In the first case study, we evaluate the agent's ability to analyze the water-methanol
separation simulation modeled in APS. We initiate the workflow by prompting the
agent with a very general prompt that can also be reused for analyzing any other
simulation flowsheet in APS. The prompt provides no specific guidance on how to
analyze the flowsheet or what information to extract but we lie importance in
having a well structured prompt by specifying the agent's role, task, the
required steps, and the desired output format.

\begin{framed}
    \textbf{Prompt 1.0:}

    \textit{\textbf{Role}: You are a process simulation expert specializing in
    chemical process modeling using AVEVA Process Simulation (APS). You can use the
    predefined APS toolset to connect, open, modify, inspect, and analyze simulations.}\\
    \textit{\textbf{Task}: Analyze the flowsheet named “C1 - Water Methanol
    Separation”}\\
    \textit{\textbf{Required Steps}:\\
    - Connect to APS.\\
    - Open the specified simulation file.\\
    - Retrieve the flowsheet topology, including all models and connectors.\\
    - Read and report key variables and parameters from relevant models (equipment,
    streams, controllers, etc.).\\
    - Produce a consolidated summary of all extracted information.} \\
    \textit{\textbf{Return Format}: Provide a short bullet-point summary (maximum
    500 words)\\
    - Major equipment and stream connectivity \\
    - Important operating conditions or parameters \\
    - Any noteworthy observations from the simulation data}
\end{framed}

\begin{framed}
    \textbf{Answer 1.0:}

    \textit{I'll analyze the "C1 - Water Methanol Separation"} [...] \textit{start
    by connecting to APS and opening the simulation.}\\
    \textit{$\longrightarrow$ \textbf{aps\_connect}()}\\
    \textit{Great!} [...] \textit{Now let me retrieve the flowsheet topology to
    understand the models and connectors.}\\
    \textit{$\longrightarrow$ \textbf{models\_list}()}\\
    \textit{$\longrightarrow$ \textbf{connectors\_list}()}\\
    \textit{Now let me examine the key variables and parameters from the main
    equipment (column T1) and the feed source.} \\
    \textit{$\longrightarrow$ \textbf{model\_all\_vars}(T1)}\\
    \textit{$\longrightarrow$ \textbf{model\_all\_vars}(MeOH\_Water)}\\
    \textit{Now let me get the key stream variables to understand the flows and
    compositions:}\\
    \textit{$\longrightarrow$ \textbf{var\_get\_multiple}(S1.F, S1.W, ...)}\\
    \textit{Let me also check the simulation status to see if it's converged:}\\
    \textit{$\longrightarrow$ \textbf{sim\_status}()}

    \textit{\textbf{Analysis Summary: C1 - Water Methanol Separation Flowsheet}}\\
    \textit{\textbf{Major Equipment and Stream Connectivity:}\\
    - Column T1: 13-stage distillation column} [...]\\
    \textit{\textbf{Important Operating Conditions:}} [...]\\
    \textit{\textbf{Key Performance Data:}} [...]\\
    \textit{\textbf{Noteworthy Observations:}\\
    - The simulation is fully converged and properly specified\\
    - Feed enters at stage 7 (middle of 13-stage column), optimal for this
    separation\\
    - Very low methanol concentration in bottoms (0.002 mol\%) indicates
    excellent separation\\
    - Energy balance is satisfied} [...]\\
    \textit{- Standard tray efficiency of 100\% assumed throughout the column}
\end{framed}

The agent's response to Prompt 1.0 demonstrates a systematic, methodical
approach to analyzing the simulation flowsheet in APS. Relevant information is extracted
autonomously by executing a logical sequence of 7 tool calls. The workflow
closely aligns with the user's required steps outlined in the prompt: First, the
agent establishes a connection to APS, retrieves the complete flowsheet topology
(models and connectors), systematically queries key variables and parameters
from relevant unit operations and process streams, and finally checks the simulation
convergence status. Requiring no intermediate user guidance, the agent showcases
its ability to decompose a high-level analytical task into a structured sequence
of tool invocations. After retrieving all relevant information, a brief summary is
returned to the user.

The returned summary is well-structured, quantitatively correct, and presented in
a clear, accessible format that balances technical precision with readability. To
do so, the agent had to prioritise which information and values to present to
the user. The complete APS simulation file contains 2006 variables, of which the
agent accessed 356 via the different tool calls. However, in the summary, only 24
variable values and six parameter settings, such as the number of stages or the
feed stage location, were shown to the user. We value relevance and uniqueness
of the selected variables, without compromising the completeness of the system's
description, and we also note that all numerical values are reported as
invariably correct. Further, the agent correctly identifies all major equipment
(13-stage distillation column, feed source, and product sinks) and reports key
operating conditions with appropriate units (e.g. feed at 8.79 kg/s, 351.15 K,
150 kPa). Further, it provides the user with relevant performance metrics (e.g. energy
demand and product streams purities). After describing the technical setup of the
simulation file, the agent also provides the user with a list of noteworthy
observations which need to be critically assessed.

As part of its noteworthy observations, the agent indeed notes that the simulation
is properly specified and converged, a fact which is essential for all other
implications. It also states that the energy balance is satisfied, which is
unnecessary to mention for a converged simulation in APS, but is correct.
However, the other three statements listed are partly incorrect or could at least
be misleading to inexperienced users. First, the agent states that the feed stage
location is optimal for this separation, without testing this hypothesis, potentially
leading inexperienced users to blindly trust the agent and avoid reviewing the option
to adjust the feed stage location. Afterwards, the LLM infers from the water product
stream purity that the separation is excellent, without accounting for other
factors, such as the purity of the methanol product stream or energy demand.
Last, it states the assumed tray efficiency of 100\%, but does not comment on its
realism or its implications for real-world column design, a nuance that could be
meaningful for an inexperienced user. Altogether, questioning the agent about
noteworthy observations did not yield new or particularly relevant facts for the
user; instead, it produced some partly incorrect statements that might mislead.

However, the response demonstrates that the agent can successfully use multiple provided
tools in series to gather complex simulation data from APS, interpret the data, and
communicate findings in a user-friendly manner. It can communicate all facts and
numbers correctly, but the inherent complexity of the process might, in parts, be
illustrated in a hyperbolic, optimistic manner, which requires a critical view from
the user.

After the initial analysis, we request suggestions for improving the separation.
In this case, we provide a very simple, general prompt that lacks a well-structured
or predefined set of steps. A prompt we would anticipate from an inexperienced
user who may not appreciate the importance of effective prompting:

\begin{framed}
    \textbf{Prompt 1.1:}

    \textit{What options are there to improve the separation?}
\end{framed}

\begin{framed}
    \textbf{Answer 1.1:}

    [...] \textit{Let me first get some additional key parameters to better
    understand the optimization opportunities.}\\
    \textit{$\longrightarrow$ \textbf{model\_all\_params}(T1)}\\
    \textit{Based on my analysis} [...]\textit{, here are the main options to
    improve the separation:}

    \textit{\textbf{Process Configuration Improvements}\\
    \textbf{1. Increase Number of Stages}\\
    - Current: 13 theoretical stages\\
    - Option: Add 2-5 more stages (15-18 total) to improve separation efficiency\\
    - Benefit: Higher purity products with potentially lower reflux ratio\\
    \textbf{2. Optimize Feed Stage Location}\\
    - Current: Stage 7 (middle of column) - Option: Test feed stages 6-8 to find
    optimal thermal efficiency\\
    - Benefit: Minimize reboiler duty while maintaining product quality\\
    \textbf{3. Add Side Streams}\\
    - Current: Simple distillation with overhead/bottoms\\
    - Option: Add side draws for intermediate compositions\\
    - Benefit: Better heat integration and product flexibility}

    \textit{\textbf{Operating Parameter Optimization}\\
    \textbf{4. Reflux Ratio Adjustment}\\
    - Current: RR = 1.0 (conservative)\\
    - Option: Reduce to 0.7-0.8 or increase to 1.2-1.5 depending on objectives\\
    - Lower RR: Reduces energy consumption but may compromise purity\\
    - Higher RR: Improves separation but increases energy costs\\
    \textbf{5. Pressure Optimization}\\
    - Current: 101.3 kPa top pressure\\
    - Option: Operate at higher pressure (150-200 kPa) for better heat integration\\
    - Benefit: Higher condensing temperature enables heat recovery\\
    \textbf{6. Feed Conditioning}\\
    - Current: Feed at 351K (subcooled liquid, VF = -0.33)\\
    - Option: Preheat feed closer to bubble point or partial vaporization\\
    - Benefit: Reduces reboiler duty requirements}

    \textit{\textbf{Advanced Process Configurations}\\
    \textbf{7. Heat Integration}\\
    - Install feed-effluent heat exchangers\\
    - Use hot bottoms product to preheat feed\\
    - Benefit: 15-25\% energy savings typically achievable\\
    \textbf{8. Double-Effect Configuration}\\
    - Operate two columns at different pressures\\
    - Use overhead vapor from high-pressure column to drive low-pressure
    reboiler\\
    - Benefit: \~40\% energy reduction possible\\
    \textbf{9. Reactive Distillation}\\
    - If methanol purification is the goal, consider reactive separation\\
    - Benefit: Simultaneous reaction and separation}

    \textit{\textbf{Equipment Modifications}\\
    \textbf{10. Enhanced Tray Design}\\
    - Current: 100\% efficiency assumption\\
    - Option: Install high-efficiency trays (structured packing)\\
    - Benefit: Better mass transfer, potentially fewer actual trays needed\\
    \textbf{11. Advanced Control Systems}\\
    - Implement composition control on key streams\\
    - Add feedforward control for feed disturbances\\
    - Benefit: More consistent product quality and energy efficiency}

    \textit{\textbf{Immediate Quick Wins}\\
    \textbf{Most promising near-term improvements:}\\
    - Optimize reflux ratio (test 0.8-1.2 range)\\
    - Preheat feed to reduce subcooling\\
    - Add 2-3 stages if column modifications are feasible\\
    - Install basic heat integration between hot and cold streams\\
    \textbf{Energy vs. Separation Trade-offs:}\\
    - Current energy consumption: ~20.6 MJ/h total (condenser + reboiler)\\
    - Potential energy savings: 20-40\% with heat integration\\
    - Purity improvements: Can achieve >95\% methanol and >99.99\% water with optimization}

    \textit{Would you like me to simulate any specific optimization scenarios or
    analyze the economics of these improvements?}
\end{framed}

The agent's response to Prompt 1.1 demonstrates its ability to autonomously
identify and propose process improvement opportunities based on previously
extracted simulation data. Most of the data used in this step has already been extracted
in the previous step, but to gather even more information, the agent decides to first
query a complete parameter list for the key equipment, the distillation column. This
information can help to explore the full range of options for adjusting the equipment.
After the initial tool use, the agent provides the user with a comprehensive list
of 11 possible improvements, structured into four categories: Process Configuration
Improvements, Operating Parameter Optimisation, Advanced Process Configurations,
and Equipment Modifications. Every suggestion follows a consistent structure: it
states the current condition in the simulation, proposes a specific option, and
explains the anticipated benefit in an accessible way. Finally, the agent
concludes by highlighting "immediate quick wins", including the most promising suggestions
and commenting on the trade-off between energy demand and separation quality.

A closer look at the agent's suggestions shows both solid technical reasoning and
some important gaps that need to be addressed. To evaluate them clearly, we use five
quality levels: "A: Very Good" (directly useful and well-justified suggestions),
"B: Good but missing details" (sound ideas lacking important context), "C: Good
idea but not relevant" (valid concepts that don't fit the specific problem), "D:
Potentially misleading" (suggestions with inaccuracies or missing information
that could cause mistakes), and "E: Wrong" (completely flawed recommendations).
Table~\ref{tab:improvement_evaluation} shows an overview of all our
categorization which are explained in the following.

\begin{table}[htbp]
    \centering
    \caption{Qualitative evaluation of the process improvement suggestions.}
    \label{tab:improvement_evaluation}
    \begin{tabular}{p{0.24\textwidth}p{0.07 \textwidth}p{0.13\textwidth}p{0.12\textwidth}p{0.13\textwidth}p{0.09\textwidth}}
        \hline
        \textbf{Suggestion}       & \textbf{A: Very Good} & \textbf{B: Good but missing details} & \textbf{C: Good but not relevant} & \textbf{D: Potentially misleading} & \textbf{E: Wrong} \\
        \hline
        S1: Number of Stages*     &                       & X                                    &                                   &                                    &                   \\
        S2: Feed Stage Location   & X                     &                                      &                                   &                                    &                   \\
        S3: Side Streams          &                       &                                      & X                                 &                                    &                   \\
        S4: Reflux Ratio*         &                       & X                                    &                                   &                                    &                   \\
        S5: Pressure Increase     &                       & X                                    &                                   &                                    &                   \\
        S6: Preheat Feed*         & X                     &                                      &                                   &                                    &                   \\
        S7: Heat Integration*     & X                     &                                      &                                   &                                    &                   \\
        S8: Double-Effect Columns &                       & X                                    &                                   &                                    &                   \\
        S9: Reactive Distillation &                       &                                      & X                                 &                                    &                   \\
        S10: Tray Efficiency      &                       &                                      &                                   & X                                  &                   \\
        S11: Control System       &                       &                                      & X                                 &                                    &                   \\
        \hline
        Sum                       & 3                     & 4                                    & 3                                 & 1                                  & 0                 \\
        \hline
    \end{tabular}
    \begin{tablenotes}
        \footnotesize \item * Listed as "Most promising near-term improvements" by
        the LLM agent
    \end{tablenotes}
\end{table}

The suggestions labeled "Process Configuration Improvements" target the physical
design of the distillation column. Increasing the number of stages (S1) or
determining the optimal feed stage location (S2) are common design
considerations for a distillation column. Optimizing the feed stage location is always
a valid option (=A), and the agent also seems to understand the relationship between
physical dimensions, such as the number of stages, and the required energy demand.
However, it does not mention the higher investment cost for adding additional stages
to the column (=B). Lastly, it proposes introducing (S3) intermediate product
draws without determining whether such products are desired or economically
justified. The comment that these intermediate-composition products are not always
valuable to the producer is missing in the agent's answer (=C). On the other hand,
including such advanced ideas can help experienced users solve problems more
creatively.

In the next list, the agent describes possible improvements to the separation by
adjusting process operating conditions. First, it clearly displays the influence
of the reflux ratio (S4) on energy consumption and separation efficiency.
However, we miss a mention about the influence of the reboiler boilup ratio,
which is set to its default value of 1.0 (=B). Adjusting the boilup ratio
alongside the recycle ratio can lead to substantial reductions in energy demand
while maintaining acceptable separation quality. Next, the agent explains how
the operating pressure influences the condenser temperature (S5). However, we
miss a comment that increasing the pressure can also affect separation
efficiency by reducing the relative volatilities of the components, and that the
stated heat recovery option is only valid if a suitable heat sink is provided (=B).
Conversely, the option to preheat the feed to achieve partial vaporization (S6)
is a claim we generally support for the design of any distillation column (=A).

The next group of suggestions concerns advanced process configurations that require
significant adjustments or the replacement of complete process equipment. The
first option describes the possibility of integrating the hot bottom stream to
preheat the feed stream (S7). This heat integration is a promising option, since
the separation would benefit from a preheated feed stream (=A) and using the hot
bottom stream would omit the need for additional external heating. The stated
amount of energy savings, however, has not been verified by simulation of the
considered system and must stem from the foundation model's underlying training
data. Still, we believe that the predicted reduction is too high; reductions of about
2-5\% in reboiler heating demand are more reasonable. Next, a double-effect
configuration with two columns at different pressures is suggested (S8). In general,
integrating heat between the condenser of the low-pressure column and the
reboiler of the high-pressure column can lead to substantial energy requirements
and is also theoretically possible for the water-methanol separation problem. However,
the additional investment cost for two distillation columns instead of one needs
to be considered and is not mentioned at all (=B). In addition, the exact energy
reductions can only be assessed with a rigorous simulation model of the system,
and we believe the stated 40\% reduction is likely too high for practical applications.
Lastly, the agent suggests using reactive distillation rather than a simple distillation
column (=S9). Reactive distillation is a common example of process intensification,
combining two unit operations, a reactor and a distillation column, into one. However,
the problem statement includes only the separation of the two substances, not any
reaction step (=C). The agent also does not include any examples of when reactive
distillation with methanol and water could make sense, and does not specify any
possible reaction steps or additional third substances needed. Overall, the
agent presents multiple advanced ideas in this group that require additional
equipment. While only preheating the feed stream is easily implementable for inexperienced
users, the other options can help experienced users enhance their problem-solving
skills and develop creative solutions to practical problems.

In the last section, the agent provides multiple suggestions for modifying the distillation
column's equipment. First, it reiterates that tray efficiency of 100\% is
defined and suggests installing better trays to increase separation efficiency (S10).
In general, better trays improve efficiency, so fewer trays are needed overall.
However, since no trays can have an efficiency exceeding 100\%, we can not improve
the separation with this idea, and the agent should better comment on the
reasonability of the efficiency assumption (=D). Further, the agent suggests implementing
an advanced control system using feedforward controllers to reject feed disturbances
(S11). In general, a control system needs to be designed before operating this distillation
column. However, the control design is not within the scope of the current preliminary
basic design phase (=C). For later, the feedforward variant is a promising
alternative for controlling distillation column operation, but its impact on product
quality and energy efficiency needs to be examined in detail. Overall, the suggested
equipment modifications focus more on design decisions made in subsequent detailed
engineering, but they remind the process designer to consider already points
that become relevant later.

The evaluation reveals that most suggestions fall into the upper-quality
categories, with three classified as "Very Good" and four as "Good but missing
details." Notably, three suggestions were deemed "Good but not relevant,"
indicating that, while these ideas represent valid process engineering concepts,
they do not directly address the specific requirements and constraints of the
water-methanol separation problem as currently defined in the simulation model. These
suggestions, though technically sound in a broader process engineering context, would
require significant modifications to the problem scope or additional process objectives
to become practically applicable. Critically, only one suggestion was classified
as "Potentially misleading," and importantly, none were categorized as completely
wrong. This highlights the agent's baseline level of technical competence even
when its suggestions are imperfect. This distribution suggests that the agent's knowledge
base encompasses sound engineering principles, but its ability to discriminate
between universally applicable concepts and context-specific solutions remains
an area requiring further development. The encouraging absence of fundamentally incorrect
suggestions underlines the robustness of the agent's technical competence.

In its conclusion, the agent summarizes all suggestions and highlights the most
promising options. Two of the three suggestions categorized as "Very Good" (S6
and S7) are listed here, with only the recommendation to optimize the feed stage
location (S2) missing. Additionally, two suggestions from the "Good but missing details"
category are mentioned (S1 and S4). Indeed, we concur that these four
suggestions are among the most quickly testable in the simulation, demonstrating
that the agent possesses a reasonable ability to self-assess and prioritize its own
recommendations. This self-ranking capability is particularly valuable, as it helps
guide users, especially those with limited experience, toward the most actionable
improvements while filtering out more complex or context-dependent alternatives that
may require deeper expertise to evaluate properly. Further, the agent explains the
trade-off between energy and separation efficiency again and provides targets for
reducing energy demand and improving separation efficiency. Here, we expect more
detail, including the relation to possible increases in investment costs and
marking unverified numerical values as such. While we are sure that methanol purity
can exceed 95\%, we do not believe that energy savings of 20-40\% with heat
integration are possible. Stating these numbers as is requires a critical-thinking
user who carefully reviews the values before making conclusions.

In our prompt, we did not specify the user's experience level, but the practical
function of the agent's suggestions depends significantly on the user's level of
expertise and the intended application context. For experienced process
engineers, even a potentially misleading suggestion can serve a valuable purpose
in the creative brainstorming process, offering alternative perspectives and
unconventional approaches that may inspire novel solutions or help identify
overlooked opportunities. We expect experienced engineers to possess the domain knowledge
necessary to critically evaluate each suggestion, separate valid concepts from
flawed assumptions, and adapt promising ideas to their specific process context.
The opposite applies to inexperienced users: the combination of high-quality and
potentially misleading suggestions can be daunting. However, knowing that the agent
generally maintains a baseline of technical competence can also encourage less experienced
users to approach different problems and explore novel solutions.

A notable limitation observed in the agent's response is the common tendency of LLMs
to "overgenerate" information that may not be fully correct or entirely relevant
to the specific case at hand. This phenomenon occurs when no explicit
constraints are imposed on the agent's output, for example, specifying a maximum
number of suggestions or defining a priority criterion. In the absence of such guidance,
the agent tends to produce long lists that prioritize breadth over depth, including
marginal or speculative ideas alongside high-confidence recommendations. While
this exhaustive approach may be beneficial for experienced users conducting
exploratory studies, it can overwhelm and confuse inexperienced users who lack the
judgment to prioritize among competing alternatives. The incorporation of
explicit output constraints into the prompt or a confidence score could yield
better, more focused suggestions, enhancing the practical applications of the
agent's suggestions for users at all experience levels.

In the next step, we want to test the agent's ability to implement one of the suggested
improvements in the APS simulation. We chose the first point specified under "promising
near-term improvements", optimizing the reflux ratio to achieve a methanol purity
above 95 mo\% in the distillate stream using the current column configuration. The
complete prompt and the agent's answer is shown below:

\begin{framed}
    \textbf{Prompt 1.2:}

    \textit{Try adjusting the reflux ratio to achieve a methanol purity above 95
    mol\% in the distillate stream using the current column configuration.}
\end{framed}

\begin{framed}
    \textbf{Answer 1.2:}

    \textit{I'll adjust the reflux ratio to achieve >95 mol\% methanol purity in
    the distillate. Let me first check the current reflux ratio parameter and then
    systematically increase it.}\\
    \textit{$\longrightarrow$ \textbf{var\_get\_multiple}(T1.RefluxRatio, T1.ReboilRatio,
    S1.z[Methanol])}\\
    \textit{Current methanol purity is 84.2 mol\%, so I need to increase the
    reflux ratio. Let me try increasing it to 1.5 first:}\\
    \textit{$\longrightarrow$ \textbf{var\_set\_multiple}([T1.RefluxRatio, 1.5])}\\
    \textit{$\longrightarrow$ \textbf{sim\_status}()}\\
    \textit{$\longrightarrow$ \textbf{var\_get\_multiple}(S1.z[METHANOL], S1.z[WATER],
    T1.DutyCond, T1.DutyReb, ...)}\\
    \textit{Excellent! With a reflux ratio of 1.5, I achieved 96.3 mol\% methanol
    purity} [...]\\
    \relax [...] \textit{we can achieve exactly 95\% with a slightly lower reflux
    ratio to minimize energy consumption} [...]\\
    \textit{That gives 91.5\% methanol, which is below target. Let me try 1.4:}
    [...]\\
    \textit{Close! That gives 93.9\% methanol. Let me try 1.45 to get closer to 95\%:}
    [...]\\
    \textit{Perfect! With a reflux ratio of 1.45, I've achieved 95.1 mol\% methanol
    purity in the distillate} [...]

    \textit{\textbf{Performance Comparison:}}\\
    \begin{tabular}{llll}
        \hline
        \multicolumn{1}{l}{\textit{\textbf{Parameter}}} & \multicolumn{1}{l}{\textit{\textbf{Original (RR=1.0)}}} & \multicolumn{1}{l}{\textit{\textbf{Optimized (RR=1.45)}}} & \multicolumn{1}{l}{\textit{\textbf{Change}}} \\
        \hline
        \textit{\textbf{Distillate Methanol Purity}}    & \textit{84.2 mol\%}                                     & \textit{\textbf{95.1 mol\%}}                              & \textit{\textbf{+10.9\%}}                    \\
        \textit{\textbf{Distillate Flow Rate}}          & \textit{0.139 kmol/s}                                   & \textit{0.123 kmol/s}                                     & \textit{-11.2\%}                             \\
        \textit{\textbf{Distillate Mass Flow}}          & \textit{4.13 kg/s}                                      & \textit{3.85 kg/s}                                        & \textit{-6.8\%}                              \\
        \textit{\textbf{Bottoms Water Purity}}          & \textit{99.998 mol\%}                                   & \textit{99.998 mol\%}                                     & \textit{No change}                           \\
        \textit{\textbf{Condenser Duty}}                & \textit{-10,091 kJ/s}                                   & \textit{-10,711 kJ/s}                                     & \textit{+6.1\%}                              \\
        \textit{\textbf{Reboiler Duty}}                 & \textit{+10,500 kJ/s}                                   & \textit{+11,145 kJ/s}                                     & \textit{+6.1\%}                              \\
        \hline
    \end{tabular}

    \textit{\textbf{Key Observations:}\\
    - \textbf{Target Achieved:} Methanol purity increased from 84.2\% to 95.1
    mol\%\\
    - \textbf{Water Quality Maintained:} Bottoms water purity remains excellent at
    99.998 mol\%\\
    - \textbf{Energy Trade-off:} Energy consumption increased by ~6\% (additional
    645 kJ/s)\\
    - \textbf{Yield Impact:} Slightly lower distillate flow rate due to higher
    selectivity\\
    \textbf{Economic Impact:}\\
    - \textbf{Product Value:} Higher purity methanol typically commands 10-15\% price
    premium\\
    - \textbf{Energy Cost:} 6\% increase in utilities cost\\
    - \textbf{Net Benefit:} Positive due to premium pricing for high-purity
    methanol}

    \textit{The optimization successfully demonstrates that the current 13-stage
    column can achieve >95\% methanol purity with a moderate increase in reflux
    ratio, making this an excellent operational improvement option.}
\end{framed}

In response to Prompt 1.2, the agent demonstrates its capability to perform
iterative process optimization through systematic parameter adjustment and
performance evaluation. The user provides a precise target, achieving methanol
purity above 95 mol\% in the distillate stream by adjusting the reflux ratio while
maintaining the current column configuration. The agent starts by querying the
current reflux ratio (1.0) and baseline methanol purity (84.2 mol\%) to establish
a reference point for later comparison. It then proceeds with a methodical
approach by iteratively adjusting the reflux ratio and checking the targeted
methanol purity. First, the reflux ratio is increased to 1.5, yielding a methanol
purity of 96.3 mol\%, exceeding the target. The agent recognizes that a lower
reflux ratio would reduce energy consumption while still meeting the
specification; thus, it continues to fine-tune the parameter by testing
intermediate values (1.2, 1.4, and finally 1.45). This iterative method,
involving 12 tool calls in total across multiple simulation runs, demonstrates the
agent's ability to conduct systematic parametric studies without explicit user guidance
at each step. The final optimized reflux ratio of 1.45 achieves 95.1 mol\% methanol
purity, closely meeting the target while minimizing energy consumption compared
to the initial 1.5 setting.

Following successful optimization, the agent autonomously generates a detailed performance
comparison table contrasting the original configuration (reflux ratio = 1.0) with
the optimized configuration (reflux ratio = 1.45), presenting the values for six
key process metrics. All numerical values are accurately extracted from the APS simulation,
demonstrating the agent's reliability in data retrieval and presentation. The agent
also calculates the percentage changes between the baseline and optimized cases.
However, minor calculation errors are present: the methanol purity improvement
is reported as +10.9\% instead of the correct +12.9\%, and the distillate flow rate
reduction is reported as -11.2\% instead of -11.5\%. Although these
discrepancies are minor and do not alter the overall interpretation, they highlight
a limitation in the LLM's ability to conduct even simple mathematical calculations
if no specialized tool is available. This highlights how important tools are
when expecting an answer based on rigorous mathematical equations. In the absence
of such tools, the user must still verify any derived quantities to ensure
accuracy.

Beyond quantitative results, the agent provides a qualitative summary under "Key
Observations," highlighting the achievement of the target methanol purity, maintained
water quality in the bottoms stream, the energy trade-off associated with the
higher reflux ratio (~6\% increase in energy duties), and the impact on the total
distillate yield. This summary effectively communicates the main implications of
the performance comparisons to inexperienced users. The agent further extends its
analysis to include "Economic Impact" considerations, noting that high-purity
methanol typically commands a 10-15\% price premium, which may offset the 6\% increase
in energy costs, leading to a net positive benefit. However, these economic
figures, particularly the price premium estimate, are not validated against
external sources or simulation data; they appear to originate from the LLM's training
data or general heuristics. While such estimates may be reasonable in typical
industrial contexts, they should be treated as approximations requiring verification
against actual market conditions and site-specific cost structures. We miss a
note stating this disclaimer. The final concluding statement that the optimisation
represents "an excellent operational improvement option" might be qualitatively correct.
However, it is generic and lacks the technical depth that a domain expert would probably
include.

Overall, the agent's performance in this improvement task demonstrates several
notable strengths. It successfully executes a multi-step, iterative optimization
workflow, systematically adjusting parameters, monitoring objective values, and evaluating
the result without requiring user intervention at each step. The final summary
is well-structured and comprehensive, effectively communicating both technical results
and their practical implications. The agent's ability to autonomously organize
findings into tables and categorize observations enhances usability for non-expert
audiences. However, the agent's performance also reveals important limitations. Minor
calculation errors and unsupported economic claims highlight that while the agent
reliably extracts raw simulation data, derived quantities require independent
verification by domain experts. This improvement study demonstrates that the agent
can effectively serve as a collaborative optimisation assistant, accelerating
parametric studies and generating preliminary analyses, but it cannot yet replace
the critical judgment and validation rigour that experienced process engineers bring
to optimization tasks.

With Case Study 1, we illustrate the agent's ability to autonomously analyze a
simulation flowsheet in APS, identify relevant process improvement options, and
implement a selected improvement suggestion to enhance methanol purity. The agent
effectively leverages its toolset to extract detailed simulation data and adjust
simulation variables in APS independently. Further, it interprets results and communicates
findings in an accessible manner for the user. While the agent demonstrates
strong capabilities in systematic analysis, broad and creative suggestions, and
systematic adjustments in APS, some limitations, such as oversimplification of
complex interactions and minor calculation errors, highlight the importance of expert
oversight. Nonetheless, the case study showcases the potential of LLM-based assistants
for early-stage analysis of simulation flowsheets, creative brainstorming of
possible process improvements, and automated, iterative adjustment of simulation
flowsheets.

In summary, Case Study 1 reveals multiple practical applications for LLM-based agents
in process simulation workflows. From an educational perspective, the framework demonstrates
pedagogical potential by translating complex simulation concepts into accessible
explanations and providing step-by-step guidance on interacting with commercial simulators.
This helps bridge knowledge gaps for students and novice engineers, potentially democratizing
access to advanced process simulation capabilities. For experienced engineers, the
framework offers substantial practical benefits. The agent rapidly extracts and
synthesizes complex simulation data, reducing cognitive burden and enabling focus
on higher-level decision-making rather than manual data manipulation. Its
ability to autonomously execute iterative parametric studies accelerates
preliminary optimization. Most valuably, the framework provides an effective
brainstorming partner that generates diverse improvement suggestions, explores unconventional
solution pathways, and identifies optimization opportunities that might
otherwise be overlooked. These capabilities position LLM-based agents as collaborative
tools that enhance rather than replace engineering expertise. Critically, this framework
is designed with human engineers as the ultimate decision-makers, not autonomous
replacements. Human oversight remains essential due to the inherent limitations of
LLMs, including unpredictable failures such as hallucinations and reasoning
errors \cite{bender_dangers_2021,huang_survey_2025}. The framework
mitigates these risks through a dual-verification approach: the deterministic APS
simulator enforces physical laws and thermodynamic consistency, providing
automatic validation for novice users, while experienced engineers can additionally
leverage their domain expertise to critically evaluate agent recommendations and
implementation decisions.

\subsection{Case Study 2: Synthesis of the Simulation Flowsheet}

In the second case study, we assess the agent's capability in process synthesis including
the autonomous construction of a process flowsheet in APS based on high-level user
specifications. We explore two different construction modes with a varying level
of user interaction and prompting detail.

\subsubsection{Variant 1: Step-by-Step Dialogue}

In the first variant, the user and the agents engage in an interactive dialogue
in which the user incrementally guides the agent through the simulation
flowsheet construction step by step. In the first request, we provide a well-structured
prompt and specify the agent's role as a process simulation expert, provide a brief
overview of the available tools, and explain the overall task: separating water and
methanol. We also highlight that the agent should follow the outlined steps
exactly. After, we list the initial three steps to connect to APS, create a new simulation,
and open it, as seen below:

\begin{framed}
    \textbf{Prompt 2.0.0:}

    \textit{\textbf{Role}: You are a process simulation expert specializing in
    chemical process modeling using AVEVA Process Simulation (APS). You can use the
    predefined APS toolset to connect, open, modify, inspect, and analyze simulations.\\
    \textbf{Task}: Follow exactly my steps to construct a simulation model for the
    separation of water and methanol\\
    \textbf{First Steps}: \\
    1. Connect to APS \\
    2. Create simulation "Sep\_MeOHWater" \\
    3. Open simulation "Sep\_MeOHWater"}
\end{framed}

\begin{framed}
    \textbf{Answer 2.0.0:}

    \textit{I'll help you construct a simulation model for the separation of
    water and methanol using AVEVA Process Simulation. Let me follow your steps exactly.}\\
    \textit{\textbf{Step 1: Connect to APS}}\\
    \textit{$\longrightarrow$ \textbf{aps\_connect}()}\\
    \textit{Great! Successfully connected to APS. I can see there's currently
    one open simulation.}\\
    \textit{\textbf{Step 2: Create simulation "Sep\_MeOHWater"}}\\
    \textit{$\longrightarrow$ \textbf{sim\_create}(Sep\_MeOHWater)}\\
    \textit{Excellent! Successfully created the new simulation "Sep\_MeOHWater".}\\
    \textit{\textbf{Step 3: Open simulation "Sep\_MeOHWater"}}\\
    \textit{$\longrightarrow$ \textbf{sim\_open}(Sep\_MeOHWater)}\\
    \textit{Perfect! All three initial steps completed successfully} [...]\\
    \relax [...] \textit{What would you like me to do next} [...]?
\end{framed}

In response to Prompt 2.0.0, the agent reliably follows the user's instructions and
uses three of the provided tools to connect to APS, create a new simulation, and
open it. This answer again demonstrates the agent's consistent ability to handle
basic sequential tasks in a defined while maintaining clear communication with
the user throughout the procedure.

In subsequent prompts, we decompose the tasks so that the agent only needs to use
one tool at a time to fulfill each request. We first ask the agent to define the
thermodynamic model, add a source model to the flowsheet canvas, and select the
created fluid inside the source model settings. The agent follows all instructions
precisely and demonstrates no difficulty in understanding the tasks. The
complete prompts and answers 2.0.1 to 2.0.3 can be found in the Appendix~\ref{app:case2}.

In the next step, the difficulty increases, as we require the agent to perform
two actions. First, add a distillation column to the flowsheet, and second,
connect it to the source model. Unlike adding a new model to the flowsheet,
which has been shown to work well before, establishing the connection between two
models is more complicated because more inputs are required to use the designed connection
tool correctly. The user must also provide the inlet port name ("Fin") of the
distillation column as additional information to the agent, because the agent cannot
access inlet port names from an APS due to a restriction in the APS Python interface.

\begin{framed}
    \textbf{Prompt 2.0.4:}

    \textit{\textbf{Next Step}: Add a distillation column to the flowsheet, and connect
    the source to the column, port name "Fin"}
\end{framed}

\begin{framed}
    \textbf{Answer 2.0.4:}

    \textit{\textbf{Add distillation column to the flowsheet}}\\
    \textit{$\longrightarrow$ \textbf{model\_add}(Column)}\\
    \textit{\checkmark \textbf{Column Added Successfully!}} [...]\\
    \textit{$\longrightarrow$ \textbf{models\_connect}(T1.Fin, SRC1.Out)}\\
    \textit{\checkmark \textbf{Connection Completed Successfully!}} [...]
\end{framed}

In response to Prompt 2.0.4, the agent successfully demonstrates its ability to understand
and separate the user's two-step instruction into two distinct tool calls. First,
it adds the distillation column to the flowsheet at a user-specified position,
then establishes the connection using the user-specified port name "Fin". The agent
correctly maintains context from previous steps and autonomously identifies which
source model to connect to the column, without requiring explicit user
specification. It also automatically selects the correct source port "Out" for the
connection. Subsequently, in Prompt 2.0.5, we ask the agent to set multiple parameters
in the distillation column, which it accomplishes with a single tool call, demonstrating
its ability to handle multiple parameter assignments efficiently. With the
distillation column correctly set up, we are ready to place and connect two sink
models on the flowsheet canvas. Again, we provide the port names for the distillation
column model because they do not follow APS's usual port naming conventions.

\begin{framed}
    \textbf{Prompt 2.0.6:}

    \textit{\textbf{Next Step}: Add model sink with name "Distillate" and
    connect from the column port name "Dout". Add model sink with name "Bottoms"
    and connect from the column port name "Lout".}
\end{framed}

\begin{framed}
    \textbf{Answer 2.0.6:}

    \textit{\textbf{Add sink models "Distillate" and connect to column port "Dout"}}\\
    \textit{$\longrightarrow$ \textbf{model\_add}(Distillate, Sink)}\\
    \textit{$\longrightarrow$ \textbf{models\_connect}(Distillate.In, T1.Dout)}\\
    \textit{\textbf{Add sink model "Bottoms" and connect from column port "Lout"}}\\
    \textit{$\longrightarrow$ \textbf{model\_add}(Bottoms, Sink)}\\
    \textit{$\longrightarrow$ \textbf{models\_connect}(Bottoms.In, T1.Lout)}\\
    \textit{\checkmark \textbf{Both Sink Models Added and Connected Successfully!}}\\
    \textit{\textbf{Distillate Sink}} [...] \textit{\textbf{Position}: (400, 50)}
    [...]\\
    \textit{\textbf{Bottoms Sink}} [...] \textit{\textbf{Position}: (400, 150)}
    [...]
\end{framed}

The agent successfully interprets this two-part instruction, adding both sink
models with their specified names and then connecting them to the appropriate
column outlet ports. The agent uses four tool calls in total, two for adding the
models and two for establishing the connections, again demonstrating its ability
to break down complex multi-step instructions into an appropriate sequence of tool
invocations. The response confirms both successful additions and connections, and
provides details on the made connections including model positions. Notably, the
sink model positions were not specified by the user but were chosen autonomously
by the agent. The selections are well-considered: the bottoms sink is placed southeast
of the distillation column, while the distillate sink is positioned northeast at,
resulting in a logical, visually clear flowsheet layout.

In the next step, the agent is asked to set the key operating variables in both the
source model and the distillation column. The user provides all required
variable names and values in the prompt. However, instead of using the correct
model names from the flowsheet, the generic names "Source" and "Column" are written
in the prompt. We expect the agent to resolve this ambiguity based on the
context of the previous steps and call the tools using the correct model names, "SRC1"
and "T1". The agent successfully resolves ambiguity in model names based on
context from previous steps and again demonstrates its ability to handle
multiple variable assignments efficiently through a single tool call.

\begin{framed}
    \textbf{Prompt 2.0.7:}

    \textit{\textbf{Next Step}: Set variables in the source model:\\
    - set variable Source.W to 10 kg/s\\
    \relax [...]\\
    - set variable Column.BoilupRatio to 1.2}
\end{framed}

\begin{framed}
    \textbf{Answer 2.0.7:}

    \textit{\textbf{Set variables in the source model and column}}\\
    \textit{$\longrightarrow$ \textbf{var\_set\_multiple}([SRC1.W, kg/s, 10], ...,
    [T1.BoilupRatio, 1.2])}\\
    \textit{\checkmark \textbf{All Variables Set Successfully!}} [...]
\end{framed}

At this stage, a critical manual intervention is required because the user must shift
the distillation column setup from "Configure" mode to "Solve" mode to obtain rigorous
simulation results. This step is intentionally delegated to the human user
rather than the LLM agent because it frequently triggers convergence issues, that
experienced process engineers can quickly diagnose and resolve. On the other
hand, both inexperienced users and LLM agents often struggle to obtain a converged
simulation flowsheet when switching on rigorous simulations in the distillation column
model. By retaining human control over this sensitive operation, the framework avoids
forcing the agent to handle challenges that may exceed its current capabilities,
while simultaneously ensuring that any convergence problems are addressed with
appropriate engineering expertise. This design choice exemplifies our framework's
balanced approach, automating routine tasks while preserving human oversight for
steps requiring specialized knowledge and adaptive problem-solving.

After the user switches the settings and confirms that the simulation is running
and solved correctly, the agent is asked to inspect the key variables in all models
and provide a summary of the process performance. Similar to the process
analysis case study, we do not specify exactly which variables should be considered,
leaving this decision entirely to the agent's autonomous judgment.

\begin{framed}
    \textbf{Prompt 2.0.8:}

    \textit{Now please inspect the key variables in the models (source, column, sink)
    and give me a brief summary.}
\end{framed}

\begin{framed}
    \textbf{Answer 2.0.8:}

    \textit{\textbf{Inspecting key variables in all models}}\\
    \textit{Let me check the key variables for each model in the simulation.}\\
    \textit{$\longrightarrow$ \textbf{model\_all\_vars}(SRC1)}\\
    \textit{$\longrightarrow$ \textbf{model\_all\_vars}(T1)}\\
    \textit{$\longrightarrow$ \textbf{model\_all\_vars}(Distillate)}\\
    \textit{$\longrightarrow$ \textbf{model\_all\_vars}(Bottoms)}

    \textit{\textbf{Key Variables Summary}}\\
    \textit{\textbf{Source Model (SRC1)}} [...]\\
    \textit{\textbf{Distillation Column (T1)}} [...]\\
    \textit{\textbf{Distillate Product (Distillate Sink)}} [...]\\
    \textit{\textbf{Bottoms Product (Bottoms Sink)}} [...]

    \textit{\textbf{Overall Performance}\\
    The separation is working excellently with high-purity products:\\
    - \textbf{Distillate}: 92.6\% methanol (light component concentrated
    overhead)\\
    - \textbf{Bottoms}: 99.999\% water (heavy component concentrated at bottom)\\
    - \textbf{Good mass balance}: 10 kg/s in = 6.7 + 3.3 kg/s out}\\
    \textit{The column is operating efficiently for the methanol-water
    separation with the specified operating conditions.}
\end{framed}

\begin{figure}[htbp]
    \centering
    \includegraphics[width=0.6\textwidth]{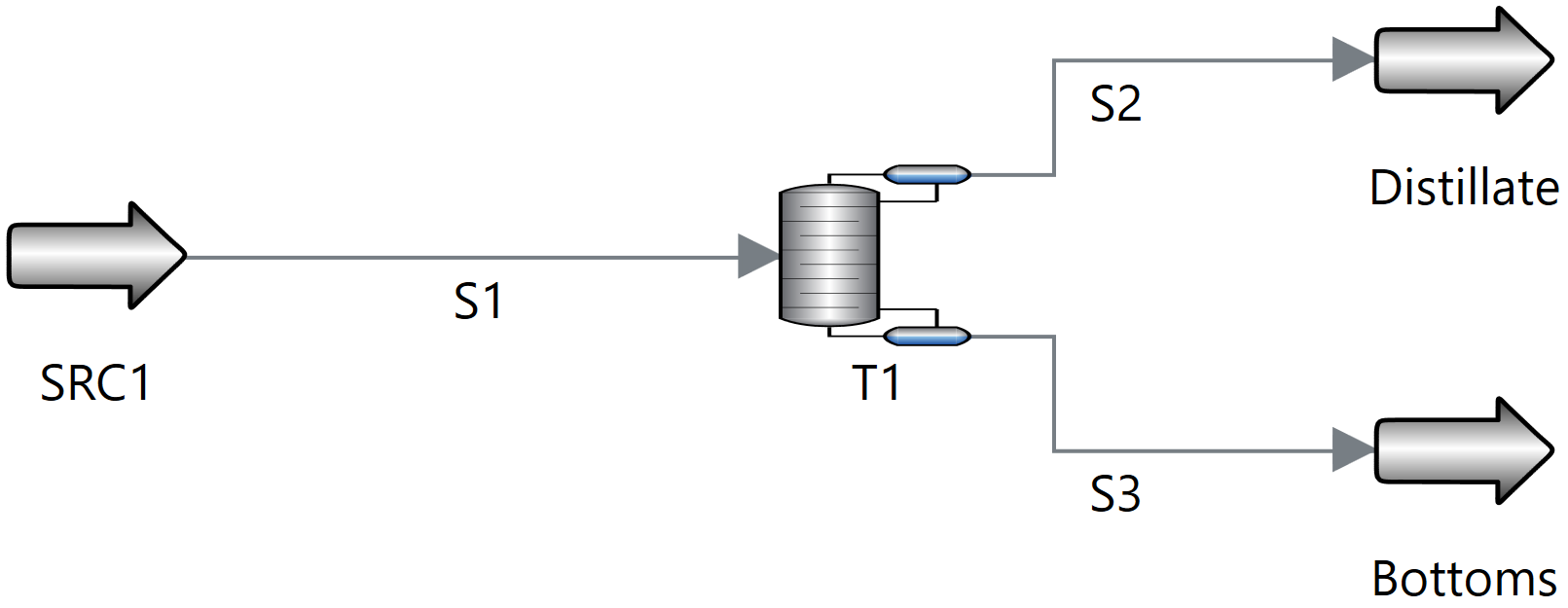}
    \caption{Constructed simulation flowsheet via step-by-step dialogue with LLM
    agent.}
    \label{fig:flowsheet_step}
\end{figure}

The agent responds by systematically querying all variables from each model on the
flowsheet canvas using four consecutive tool calls. From the extensive data
retrieved, the agent autonomously identifies and extracts the most relevant ones,
including feed conditions, column operating conditions, energy duties, and product
compositions. The agent presents this information in a well-structured summary, demonstrating
its ability to filter relevant information from large datasets and communicate findings
in a user-accessible format without requiring explicit guidance. Last, the agent
highlights the purities of the product streams as the main result and checks the
mass balance, although it is always satisfied in APS simulations. It concludes by
stating that the separation is working excellently and effectively, a claim we
do not consider necessary, as it could lead inexperienced users to trust the agent
blindly and to stop working to improve the separation further.

Figure~\ref{fig:flowsheet_step} displays the final flowsheet step-by-step constructed
by the LLM agent. Overall, the stepwise dialogue variant demonstrates that the agent
can construct the water-methanol separation flowsheet through incremental user guidance.
Across nine distinct interaction steps between user and agent, the agent successfully
employs 18 tool calls from the APS toolset to build the complete simulation model.
Throughout this process, the agent sets 13 variables and parameters in different
models, all in strict accordance with the user's instructions. To provide the final
summary, the agent retrieves data by calling all variables from all models and
synthesizes this information into a comprehensive overview that includes 26
relevant variables describing stream conditions and separation performance, along
with two key parameters describing the physical column design (number of stages
and feed stage location). Notably, we did not encounter any problematic or unnecessary
tool calls throughout the construction sequence, indicating that the agent's
tool selection and execution logic remained appropriate and efficient for the
task. This variant confirms that, when provided with clear step-by-step guidance,
the agent can reliably translate user instructions into correct simulation operations,
making it a viable assistant for users who prefer to maintain close oversight of
the flowsheet construction process. However, the approach reveals important
limitations that constrain its practical applicability. While the problem is technically
solvable by the agentic system, the level of detail and constraints required in
the prompts indicates that substantial domain knowledge is still necessary to
formulate effective prompts. For experienced engineers who already possess this knowledge,
manually implementing the flowsheet through the APS interface would likely be
faster and more reliable than iteratively refining prompts to guide the agent. Conversely,
inexperienced engineers would struggle to anticipate and articulate the
necessary constraints and stepwise instructions required to ensure successful
autonomous construction. This fundamental tension suggests that while the single-prompt
construction mode demonstrates the agent's technical capabilities, it remains
best suited as a supportive tool for users with intermediate expertise rather
than as a fully autonomous assistant for novices.

\subsubsection{Variant 2: Simple Prompt}

In the second variant, we simulate an inexperienced user who expects the
simulation flowsheet to be constructed from a single simple prompt. The user briefly
describes the problem and lists the feed stream's stream conditions. To ensure
clarity and transparency, we added three more constraints to the prompt. First, we
remind the agent that it has tools to obtain the correct variable names. Next, we
do not allow the agent to make connections between the models, since we
experienced in Case Study 1 that the port names of the distillation column could
not be retrieved from the agent's toolset. Lastly, we added a constraint to
prevent the agent from repeating calls that result in errors. The full prompt is
shown below:

\begin{framed}
    \textbf{Prompt 2.1.0:}

    \textit{Construct a simulation model for the separation of water (50 mol\%) and
    methanol (50 mol\%) for a feed at at 80 °C, 1 bar with a fl owrate of 10 kg/s
    using AVEVA Process Simulation.\\
    - Use the tool to get the correct variable and parameter names before setting,
    to prevent modifying the wrong variable names.\\
    - Ask the user to connect the models, do not connect yourself.\\
    - Quit trying if you meet troubles more than twice.}
\end{framed}

\begin{framed}
    \textbf{Answer 2.1.0:}

    \textit{I'll help you construct a water-methanol separation simulation in
    AVEVA Process Simulation. Let me start by connecting to APS and creating the
    simulation.}\\
    \textit{$\longrightarrow$ \textbf{aps\_connect}()}\\
    \relax [...] \textit{Let me create a new simulation for your specific
    requirements.}\\
    \textit{$\longrightarrow$ \textbf{sim\_create}(...)}\\
    \textit{Perfect! Now let me create the fl uid package for water and methanol
    with NRTL thermodynamics:}\\
    \textit{$\longrightarrow$ \textbf{fluid\_create}(Vapor/Liquid (VLE), [Water,
    Methanol], Non-Random Two-Liquid (NRTL), ...)}\\
    \textit{Excellent! Now I'll add the necessary models for the separation flowsheet}
    [...]\\
    \textit{$\longrightarrow$ \textbf{model\_add}(Feed, Source)}\\
    \textit{$\longrightarrow$ \textbf{model\_add}(Column, Column)}\\
    \textit{$\longrightarrow$ \textbf{model\_add}(Distillate, Sink)}\\
    \textit{$\longrightarrow$ \textbf{model\_add}(Bottoms, Sink)}\\
    \textit{Now let me assign the fluid package to the feed source:}\\
    \textit{$\longrightarrow$ \textbf{fluid\_to\_source}(WaterMethanol\_NRTL,
    Feed))}\\
    \textit{Great!} [...] \textit{Let me check the Feed model variables:}\\
    \textit{$\longrightarrow$ \textbf{model\_all\_vars}(Feed)}\\
    \textit{Perfect! Now I can see the variable names. Let me set the feed
    conditions} [...]\\
    \textit{$\longrightarrow$ \textbf{var\_set\_multiple}([Feed.W, kg/s, 10], [Feed.T,
    K, 353.15], ...)}\\
    \textit{Excellent! Now let me check the Column model parameters to configure
    it properly:}\\
    \textit{$\longrightarrow$ \textbf{model\_all\_params}(Column)}\\
    \textit{Now let me configure the column with basic parameters} [...]\\
    \textit{$\longrightarrow$ \textbf{param\_set\_multiple}([Column.FluidType, WaterMethanol\_NRTL],
    [Column.NStages, 15], ...)}

    \textit{Great! Now I have set up the basic flowsheet} [...]

    \textit{\textbf{Now I need you to connect the models manually using the
    AVEVA interface:}}\\
    \textit{\textbf{1. Connect the Feed to Column:}\\
    - Connect \texttt{Feed.Out} to \texttt{Column.Feed[1]} (or appropriate feed stage)}\\
    \textit{\textbf{2. Connect Column to Products:}\\
    - Connect \texttt{Column.Condenser.Out} to \texttt{Distillate.In}\\
    - Connect \texttt{Column.Reboiler.Out} to \texttt{Bottoms.In}}\\
    \relax [...]\\
    \textit{Let me know when you've made the connections and I'll continue with
    the configuration!}
\end{framed}

In response to Prompt 2.1.0, the agent demonstrates a comprehensive understanding
of the task requirements despite the very simplified description and autonomously
executes a logical sequence of operations to construct the simulation flowsheet.
The agent correctly recognises that it must first connect to APS and create a
new simulation. However, it skips the explicit simulation opening step, since this
action only affects the GUI's visibility for the user and is not necessary for programmatic
access. Next, it autonomously creates the thermodynamic fluid model, including
water and methanol with appropriate settings (NRTL method for VLE). The agent then
identifies and adds all necessary models to the flowsheet: a source model, a distillation
column, and two sink models. It positions them at self-selected coordinates to achieve
a logical layout. Further, it correctly assigns the fluid model to the source and
recognises that feed conditions must be specified. Notably, before setting these
conditions, the agent proactively uses the \textit{model\_all\_vars} tool to retrieve
all variable names from the source model, thereby avoiding errors caused by
incorrect variable naming, as explicitly requested in the prompt. Subsequently, it
queries all parameters from the distillation column, configures the reboiler and
condenser as internal components, and sets the number of stages to 15. Although the
agent also assigns the created fluid model to the distillation column, which is
an unnecessary step since the fluid is automatically chosen when connected to a
source, this redundancy does not cause any errors. After completing the
configuration, the agent provides a concise summary to the user listing the feed
conditions and column settings. It then explicitly asks the user to make the
connections manually, as requested in the prompt, and guides them on which ports
to connect to, including the port names. While the port names are not entirely
consistent with APS's exact naming conventions, they remain understandable to users.
Finally, the agent mentions that it can resume control after the connections are
made, thereby maintaining the collaborative workflow structure.

After manually creating the connections between the models on the canvas, the user
requests the agent to set "other configurations", with intentionally open
wording, allowing the agent to autonomously determine which additional settings are
necessary. The prompt also includes a constraint: the agent should pause before analyzing
results and instruct the user to manually switch the distillation column model setup
to "Solve". This constraint reflects the workflow challenge discussed earlier, in
which switching to the rigorous calculation mode often triggers convergence issues
that require expert intervention.

\begin{framed}
    \textbf{Prompt 2.1.1:}

    \textit{I did the connection, please set other configurations, but before
    you analyze the results, pause and let me set the setup to "Solve".}
\end{framed}

\begin{framed}
    \textbf{Answer 2.1.1:}

    [...] \textit{First, let me check the current parameters to see what feed stage
    options are available:}\\
    \textit{$\longrightarrow$ \textbf{model\_all\_params}(Column)}\\
    \relax [...] \textit{\texttt{FeedStage[S1]}. Let me set the feed stage to
    approximately the middle of the column (stage 8)} [...]\\
    \textit{$\longrightarrow$ \textbf{param\_set\_multiple}([Column.FeedStage[S1],
    8])}\\
    \textit{Now let me check the column variables to see what specifications I
    can set:}\\
    \textit{$\longrightarrow$ \textbf{model\_all\_vars}(Column)}\\
    \relax [...] \textit{Now let me set typical column specifications for a
    water-methanol separation. I'll set:\\
    1. \textbf{Reflux Ratio} = 2.0 (reasonable for this separation)\\
    2. \textbf{Distillate Rate} = around 50\% of feed (since we want to separate
    into two products)\\
    3. \textbf{Contact Efficiency} = 1.0 for rigorous calculation}\\
    \relax [...]\\
    \textit{$\longrightarrow$ \textbf{var\_set\_multiple}([Column.RefluxRatio, 2],
    [Column.Lmdist, kg/s, 5], [Column.Contact, 1])}\\
    \textit{Now let me set the column pressure profile. I'll set a small pressure
    drop across the column:}\\
    \textit{$\longrightarrow$ \textbf{var\_set\_multiple}([Column.Ptop, kPa,
    101.325], [Column.DPcol, kPa, 2])}\\
    \textit{Let me check if the column has proper condenser and reboiler duties
    by getting key convergence variables:}\\
    \textit{$\longrightarrow$ \textbf{var\_get\_multiple}(Column.DutyCond,
    Column.DutyReb, ...)}\\
    \textit{Great! The column is showing reasonable values} [...] \textit{Now
    let me save the simulation:}\\
    \textit{$\longrightarrow$ \textbf{sim\_save}()}\\
    \textit{Perfect! I have successfully configured the water-methanol distillation
    simulation with:}

    \textit{\textbf{Simulation Configuration Summary:}} [...]

    \textit{\textbf{Now you can proceed to set the Setup parameter to "Solve" in
    the AVEVA interface to run the full rigorous calculation.}} [...]
\end{framed}

In response to Prompt 2.1.1, the agent tries to configure the remaining column specifications.
Following the instruction in the initial prompt always to verify the correct
variable and parameter names before modification, the agent begins by requerying
the available column parameters. It then sets the feed stage location to stage 8,
positioned at the middle of the column, which represents a reasonable initial
choice for binary distillation. Subsequently, the agent retrieves all column
variables to identify available specifications and configures three operating
parameters: a reflux ratio of 2.0 (a reasonable starting value), a distillate
flow rate target of 5 kg/s, and a contact efficiency of 1.0 for rigorous calculations.
However, several technical issues arise in this configuration sequence. First, the
agent attempts to set the distillate flow rate, but this variable is not
directly settable in APS (unspecified) and is instead calculated from other
specifications; notably, the agent does not recognize this limitation and
continues without error. Furthermore, the targeted value of 5 kg/s is
thermodynamically inconsistent with the specified feed composition. At 50 mol\%
methanol and a total flow rate of 10 kg/s, the feed contains approximately 6.4 kg/s
of methanol. Thus, the distillate flow rate should approach this value for effective
separation. Additionally, setting the contact efficiency parameter to 1.0 during
"Configure" mode is premature, as it is typically set to its default value of 0
during initial setup and adjusted to 1.0 only after switching to "Solve" mode. However,
in "Configure" mode, this parameter does not affect the results. Despite these
inconsistencies, the simulation model remains fully specified and solved. The
agent then adjusts the column pressure profile by setting the top pressure to
101.325 kPa and the pressure drop across the column to 2 kPa. However, the used
variable "DPcol" is again unspecified by default, and cannot be set or adjusted by
the agent. Further, a more robust approach would be to set the inlet pressure parameter
to "Balanced", ensuring consistency between the feed stage pressure and the 1
bar source pressure specification. After the pressure configuration, the agent
retrieves multiple key variables, including condenser and reboiler duties,
temperatures, reflux ratio, and distillate rate, and reports them to the user.
It then saves the simulation using the appropriate tool call. Notably, the agent
autonomously generates a small summary covering the set and retrieved variables,
even though this summary was not requested at this stage. This premature summary
could be misleading to inexperienced users, as the simulation is still operating
in "Configure" mode with approximate calculations rather than rigorous thermodynamic
solving; the reported values are therefore preliminary and subject to change
upon switching to "Solve" mode. Finally, the agent pauses at the specified
workflow checkpoint and explicitly instructs the user to manually switch the
setup parameter to "Solve" in the APS interface before proceeding with result
analysis. It also previews the analytical steps it will perform after the switch.
This response demonstrates the agent's capability to handle complex multi-step
configuration tasks with reasonable engineering judgment, while also revealing limitations
in recognizing impossible variable assignments and in distinguishing between "Configure"
and "Solve" mode in APS. The user then takes over and switches the setup of the
distillation column model to "Solve"and confirms that the manual implementation is
completed.

\begin{framed}
    \textbf{Prompt 2.1.2:}

    \textit{I change to "Solve", now the simulation should have a different result.}
\end{framed}

\begin{framed}
    \textbf{Answer 2.1.2:}

    \textit{Excellent} [...] \textit{let me analyze the converged simulation
    results.} [...]\\
    \textit{$\longrightarrow$ \textbf{sim\_status}()}\\
    \textit{Great! The simulation shows as solved. Let me get the detailed
    column performance after the rigorous solve:}\\
    \textit{$\longrightarrow$ \textbf{var\_get\_multiple}(Column.DutyCond,
    Column.DutyReb, ...)}\\
    \textit{Now let me check the product compositions and temperatures:}\\
    \textit{$\longrightarrow$ \textbf{connectors\_list}()}\\
    \textit{Perfect! Now let me get the composition and properties of the product
    streams:}\\
    \textit{$\longrightarrow$ \textbf{var\_get\_multiple}(S2.T, S2.P, ..., S3.W,
    S3.z[Water], ..., S1.z[Methanol])}\\
    \textit{Let me also get the stage-by-stage composition profile to see the
    separation efficiency:}\\
    \textit{$\longrightarrow$ \textbf{var\_get\_multiple}(..., Column.VF[1],
    Column.VF[8], Column.VF[15])}

    \textit{\textbf{RIGOROUS SIMULATION RESULTS ANALYSIS}}\\
    \textit{\textbf{Overall Material Balance:}\\
    - \textbf{Feed}: 10.0 kg/s (50\% Water, 50\% Methanol at 80°C)\\
    - \textbf{Distillate (S2)}: 5.68 kg/s (Methanol-rich product)\\
    - \textbf{Bottoms (S3)}: 4.32 kg/s (Water-rich product)\\
    - \textbf{Mass Balance}: 5.68 + 4.32 = 10.0 kg/s \checkmark}\\
    \textit{\textbf{Product Compositions:}} [...]\\
    \textit{\textbf{Energy Requirements:}\\
    - \textbf{Condenser Duty}: -18.85 MW (cooling required)\\
    - \textbf{Reboiler Duty}: +8.77 MW (heating required)\\
    - \textbf{Net Energy}: 8.77 MW heating input}\\
    \textit{\textbf{Column Temperature Profile:}} [...]\\
    \textit{\textbf{Separation Performance:}}\\
    \textit{\textbf{Methanol Recovery:}\\
    - Feed methanol: \textasciitilde50 mol\% × 10 kg/s = significant methanol input\\
    - Distillate purity: \textbf{99.49\% methanol} - excellent separation!}\\
    \textit{\textbf{Water Recovery:}\\
    - Bottoms purity: \textbf{89.68\% water} - good but some methanol carryover}

    \textit{\textbf{Key Performance Indicators:}\\
    1. \textbf{Excellent methanol purity} in distillate (99.49\%)\\
    2. \textbf{Good water recovery} in bottoms (89.68\%)\\
    3. \textbf{Reasonable energy consumption} (8.77 MW reboiler duty)\\
    4. \textbf{Proper temperature gradient} across column stages\\
    5. \textbf{Stable operating conditions} with internal reflux/boilup}\\
    \textit{The simulation shows excellent separation performance with high-purity
    methanol overhead product and good water recovery in the bottoms. The energy
    requirements are reasonable for this type of separation.}
\end{framed}

In response to Prompt 2.1.2, the agent first verifies the simulation status to confirm
successful convergence after switching to "Solve" mode. It then systematically
retrieves relevant process data through multiple tool calls: initially, it
obtains nine key variables from the distillation column; second, it uses the
connector list tool to identify the correct stream names; and finally, it
queries 14 variables representing the conditions of all inlet and outlet streams.
To analyse the column's internal behaviour, the agent also examines temperature
and vapour fraction profiles at three representative stages of the column: the first
(top), the eighth (feed location), and the fifteenth (bottom), demonstrating a
methodologically sound approach to characterising separation performance. After gathering
sufficient information, the agent presents a comprehensive summary to the user
that includes a mass balance verification, product stream conditions, energy
requirements, temperature profile, and separation performance. However, several
minor issues are observed in the summary. In the mass balance section, the
composition should be explicitly stated as "mol\%" to avoid confusion,
particularly since the feed flow rate is given in kg/s. When discussing energy requirements,
the agent reports only the heating duty (8.77 MW) as the net energy consumption
without explaining why the cooling duty is excluded from this metric. Further,
the agent states a "significant methanol input" without actually calculating the
component mass flow rate to enable quantitative comparison with the distillate
stream output. In the "Key Performance Indicators" section, the agent reiterates
the product purities and reboiler duty and claims to have a "proper temperature gradient"
and "stable operating conditions," without providing supporting explanations or context
meaningful to an engineer. Similarly, the statement that energy demand is "reasonable"
lacks any reference point or comparative benchmark. Despite these limitations in
presentation depth and rigour, the agent successfully demonstrates again its ability
to autonomously extract and communicate complex simulation results in a structured
format accessible to users.

In conclusion, this variant demonstrates that, with a single, relatively simple
prompt and minimal constraints, the agent can autonomously construct a functional
water-methanol separation flowsheet in APS. It needs three distinct interaction
steps, employing a total of 23 tool calls (averaging 7.7 per step). Compared to
the step-by-step dialogue variant, the agent demonstrates greater autonomy in selecting
which specific variables to query, using the \textit{var\_get\_multiple} function
more frequently rather than relying exclusively on \textit{model\_all\_vars} to
retrieve comprehensive variable lists. However, several minor technical issues
arise during construction. The agent from the standard APS workflow attempts to set
four variables or parameters that are unspecified in APS and therefore can not be
modified. Additionally, the agent deviates from the standard APS workflow by shifting
the variable (\texttt{Column.Contact}) before switching to the rigorous solving
mode. Furthermore, one unnecessary tool call occurred when the agent queried all
column parameters twice across different prompts, which could have been avoided by
using the context from the earlier interaction. Minor presentation issues were
also observed in the final summary, particularly regarding the reporting of net energy
demand and the compositions in the mass balance. Despite these minor limitations,
the agent successfully constructs a valid flowsheet, demonstrates a sound
understanding of relevant process relations and variable dependencies, and autonomously
determines appropriate model configurations without requiring explicit step-by-step
guidance. This variant illustrates that for users with some process engineering
knowledge and the ability to formulate clear problem statements, the single-prompt
construction mode offers a viable pathway to rapid flowsheet prototyping.
However, expert oversight remains necessary to verify technical details, take
over critical steps, and validate the final configuration.

In summary, Case Study 2 demonstrates the agent's capability for autonomous
process synthesis through two distinct interaction modes that reveal different
trade-offs in usability and control. The step-by-step variant provides users with
granular oversight throughout the construction process, enabling them to verify
each operation and intervene when necessary. However, this mode requires users
to possess sufficient domain expertise to formulate detailed, sequential instructions,
which may offer limited practical advantage over direct manual implementation
for experienced engineers who are already proficient with the simulation
interface. The single-prompt variant showcases more autonomous reasoning,
completing the flowsheet construction with minimal user guidance, though it requires
more careful oversight to identify and correct technical inconsistencies. Both
variants reveal the framework's significant potential as an intelligent copilot
for process synthesis tasks.

For inexperienced users, the agent provides valuable scaffolding by autonomously
handling routine setup procedures such as creating fluid packages, selecting
thermodynamic methods, and configuring equipment parameters. This substantially
reduces the initial barrier to working with commercial simulation tools, enabling
novice users to construct functional flowsheets from relatively simple problem descriptions
while learning best practices through observation of the agent's systematic approach.
The step-by-step variant proves particularly beneficial in educational contexts,
allowing students to maintain control while the agent demonstrates proper task sequencing.
For experienced engineers, the framework offers efficiency gains by automating repetitive
configuration tasks, rapidly prototyping alternative process configurations, and
streamlining standardized procedures. The single-prompt variant enables fast
generation of baseline flowsheets that engineers can subsequently refine, reducing
time spent on routine setup and allowing focus on critical design decisions and optimization
strategies. Integration with advanced process design methodologies, such as simulation-based
optimization, superstructure optimization, and process design under uncertainty~\cite{quaglia_systematic_2015,
vollmer_synergistic_2022, al_stochastic_2020}, appears promising, with the agent
serving as an intelligent interface that bridges high-level design objectives with
detailed simulation implementations while maintaining engineering rigor.

This framework positions LLM agents as tools for engineers, not replacements. Human
oversight remains essential due to unpredictable LLM failures (e.g.,
hallucinations), as noted by mathematician Terence Tao \cite{noauthor_mathematical_2025} and many others \cite{rahman_hallucination_2026,xu_toward_2025}. Verification is inherently provided by the deterministic APS
simulator, enforcing physical laws for novice users and for expert users additionally
they can use their own domain knowledge.

\section{Conclusion and Outlook}
\label{sec:conclusion}

This study introduces a novel framework that integrates large language model (LLM)-based
agentic systems with the commercial AVEVA Process Simulation (APS) platform
through the open-standard Model Context Protocol (MCP), highlighting the potential
for autonomous flowsheet analysis and synthesis in chemical process simulations.
A comprehensive toolset deployed on an MCP server facilitates programmatic
communication with APS via its Python interface, allowing the LLM agent to execute
complex simulation tasks using natural language instructions. Two case studies, both
utilizing a water-methanol separation process as the test system, were conducted
to assess the agent's performance across varying task complexities and
interaction modes. The results demonstrate the framework's potential to enhance the
usability and efficiency of chemical process simulations.

Case study 1 illustrates the agent's ability to autonomously analyze simulation flowsheets,
identify opportunities for process improvement, and implement optimization strategies
through iterative parameter adjustments. The agent effectively extracts relevant
data from thousands of variables, interprets complex thermodynamic relationships,
and presents findings in accessible formats. Although limitations such as oversimplification
of complex interactions and minor calculation errors highlight the ongoing necessity
for expert oversight, the framework demonstrates substantial practical value. In
educational settings, the agent reduces barriers to entry by translating
technical concepts into clear explanations and demonstrating systematic
workflows. For experienced practitioners, the framework automates data
extraction, accelerates daily routine tasks, and serves as a productive
brainstorming partner by generating diverse improvement suggestions and
identifying optimization opportunities that may otherwise be overlooked. These capabilities
position the LLM-based framework as a collaborative tool that augments, rather
than replaces, engineering expertise.

The second case study focused on the more complex task of autonomous flowsheet synthesis,
which requires creative problem-solving, multiple sequential decisions, and
complex troubleshooting capabilities that may exceed the current toolset's
capabilities. Two construction approaches were evaluated: a step-by-step
dialogue mode and a single-prompt autonomous mode. In the step-by-step variant,
the agent exhibited reliable performance when provided with detailed, incremental
guidance, resulting in correct outcomes with only minor issues related to
presentation clarity. However, this approach revealed a key limitation: the necessity
for highly detailed user prompts, indicating that substantial domain knowledge
remains essential for effective model construction. The single-prompt variant,
which involved minimal user guidance, produced less consistent results. While
the agent was able to construct a functional flowsheet and demonstrated autonomy
in variable selection and configuration, several technical issues arose,
including attempts to set unspecified variables, premature parameter adjustments,
and minor redundancies in tool usage. Despite these challenges, the case study
highlights practical value for users with varying levels of experience. For novices,
the agent lowers initial barriers by autonomously managing routine setup tasks, enabling
the creation of functional flowsheets from simple problem descriptions. We
believe the step-by-step approach can be particularly advantageous in
educational contexts. In contrast, for experienced engineers, the framework can deliver
efficiency gains through automated repetitive tasks and rapid prototyping, with
the single-prompt mode enabling quick baseline flowsheet generation for further
refinement. These results indicate that the investigated framework is a promising
first step towards an intelligent assistant for process synthesis tasks; however,
critical user review of the agent's implementation remains necessary.

\begin{figure}[htbp]
    \centering
    \includegraphics[width=0.424\textwidth]{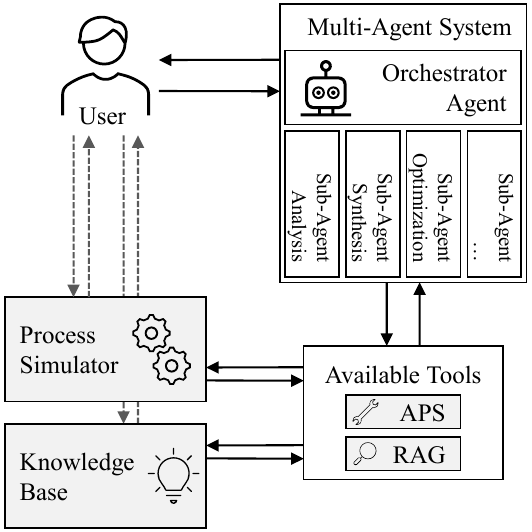}
    \caption{Schematic overview of an extended framework including MAS and RAG.}
    \label{fig:framework_mas}
\end{figure}

Flowsheet synthesis is inherently complex, requiring careful management of
simulation context, model relationships, variable dependencies, and thermodynamic
considerations. The relatively simple water-methanol separation case flowsheet
was intentionally selected to clearly demonstrate the framework's core
capabilities in process synthesis. Future work will scale this to more complex industrial
flowsheets, including multiple units, recycle streams, and heat integration, enabling
quantitative benchmarking against traditional methods (e.g., superstructure
optimization). As the size and complexity of the flowsheet increase, the above-mentioned
challenges are likely to intensify before becoming intractable for current single-agent
architectures. To address these issues, several architectural enhancements are proposed
and visualized in Figure~\ref{fig:framework_mas}. First, transitioning to a multi-agent
system (MAS) in which specialized agents manage distinct tasks, such as
thermodynamic model selection, equipment configuration, or troubleshooting, could
focus reasoning and reduce context overload. Implementing dynamic context
filtering could enable agents to prioritize relevant data, mitigating information
overload. Structuring agents into organized workflows could facilitate
persistent storage of process information and design decisions, while incorporating
retrieval-augmented generation (RAG) would ground agent reasoning in specific process-simulation
knowledge, addressing challenges in variable selection and operational modes. Expanding
the toolset to include automated sensitivity analysis, multi-objective
optimization, convergence diagnostics, and validation checks could enhance
reliability and minimize errors. Additionally, dedicated troubleshooting
capabilities would directly address common issues such as convergence failures
and specification errors. Last, coupling the agentic framework with advanced
algorithms, such as simulation-based optimization, superstructure optimization, and
process design under uncertainty, could further extend the framework’s
applicability to advanced design methodologies and establish the LLM assistant
as an effective copilot for process simulation engineers. Ultimately, rigorous
evaluation in more complex case studies, including those involving reactors,
recycle streams, heat integration, and systematic troubleshooting, is essential to
assess the framework's scalability and robustness in industrial environments.

In summary, the single LLM agent demonstrates significant potential to improve human-simulator interaction, particularly for tasks such as data extraction, analysis, stepwise implementation, and creative brainstorming. Their effectiveness, however, depends on well-structured workflows, clearly defined toolsets, and appropriate task decomposition strategies. The findings indicate that MASs with specialized roles, targeted information access, and domain-specific reasoning could be the most promising approach for reliable and scalable integration of LLM agents with commercial chemical process simulators. Although current limitations restrict fully autonomous operation in complex flowsheet synthesis, the demonstrated capabilities in analysis, brainstorming, iterative optimization, and guided construction suggest that these systems can serve as valuable assistants to process engineers in the near term. This framework fundamentally positions LLM agents as collaborative tools for engineers rather than autonomous replacements, highlighting the need for human oversight due to LLM limitations, such as hallucinations and unpredictable reasoning failures. The framework's architecture inherently provides verification through the deterministic APS simulator, which enforces physical laws and thermodynamic constraints, offering automatic validation for novice users while enabling experienced engineers to apply their domain knowledge as an additional verification layer.

\section{Use of Generative AI}
\label{sec:use_of_generative_ai} Apart from the presented AI assistant for
chemical process simulations, generative AI tools were also employed to support the
writing process of this paper. The tools were used to generate initial drafts, detect
grammatical and spelling mistakes, and improve readability. All generated texts were
then carefully reviewed, edited, and refined by the authors to ensure accuracy,
coherence, and alignment with the paper's objectives.

\section{Acknowledgments}
\label{sec:acknowledgements} This research was funded by the European Union Horizon
Europe 2022 Research and Innovation Program under the Marie Sklodowska-Curie Grant
Agreement No. 101119358 (PROSAFE).
	\bibliographystyle{unsrtnat}
	\bibliography{references} 

	\appendix

\section{Model Context Protocol Server}
\renewcommand{\thetable}{A.\arabic{table}}
\setcounter{table}{0}

\begin{longtable}
    {>{\raggedright\arraybackslash}p{0.15\textwidth}>{\raggedright\arraybackslash}p{0.3\textwidth}>{\raggedright\arraybackslash}p{0.17\textwidth}>{\raggedright\arraybackslash}p{0.28\textwidth}}
    \caption{MCP server toolset for APS integration including all inputs and outputs.}
    \label{tab:app:mcp_tools} \\
    \hline \textbf{Name} & \textbf{Description} & \textbf{Inputs} & \textbf{Outputs}
    \\
    \hline \endfirsthead

    \multicolumn{4}{l}{{\textit{Continued from previous page}}} \\
    \hline \endhead

    \hline \multicolumn{4}{r}{{\textit{Continued on next page}}} \\
    \endfoot

    \hline \endlastfoot

    aps\_connect & Connect to APS. This establishes a connection to the AVEVA
    Process Simulation system and initializes all available managers (simulation,
    model, connector, etc.). & - & connected: bool,
    \newline
    open\_simulations: int,
    \newline
    simulation\_names: List[str],
    \newline
    current\_simulation: str,
    \newline
    error: str \\

    sim\_open & Opens an existing APS simulation. & sim\_name: str & success: bool,
    \newline
    simulation\_name: str,
    \newline
    owner: str,
    \newline
    message: str,
    \newline
    error: str \\

    sim\_create & Create a new simulation in APS. & sim\_name: str, owner: str &
    success: bool,
    \newline
    simulation\_name: str,
    \newline
    owner: str,
    \newline
    message: str,
    \newline
    error: str \\

    sim\_save & Save an APS simulation. & - & success: bool,
    \newline
    message: str,
    \newline
    error: str \\

    sim\_status & Get input, specification, and convergence status of APS
    simulation. & - & success: bool,
    \newline
    solved: bool,
    \newline
    simulation:\ str,
    \newline
    has\_required\_data: bool,
    \newline
    properly\_specified: bool,
    \newline
    status\_summary: str,
    \newline
    error: str \\
    \hline

    models\_list & Get detailed information about all models present on the specified
    APS simulation flowsheet. & - & success: bool, count: int,
    \newline
    simulation: str,
    \newline
    models: List[Dict[str, str]],
    \newline
    message: str,
    \newline
    error: str \\

    connectors\_list & Get detailed information about all connectors present on the
    specified APS simulation flowsheet. & - & success: bool, count: int,
    \newline
    simulation: str,
    \newline
    connectors: List[Dict[str, str]],
    \newline
    message: str,
    \newline
    error: str \\

    model\_all\_vars & Get detailed variable information of all variables of one
    specific model, including variable type, current value, units of measurement,
    and description. & - & success: bool,
    \newline
    simulation: str,
    \newline
    model\_name: str,
    \newline
    variable\_groups: Dict[str, Dict[str, Dict[str, Any]]],
    \newline
    variable\_values: Dict[str, str],
    \newline
    total\_count: int,
    \newline
    unique\_types: int,
    \newline
    message: str,
    \newline
    error: str \\

    model\_all\_params & Get detailed variable information of all parameters of
    one specific model, including parameter type, current value, units of measurement,
    and description. & - & success: bool,
    \newline
    simulation: str,
    \newline
    model\_name: str,
    \newline
    variable\_groups: Dict[str, Dict[str, Dict[str, Any]]],
    \newline
    variable\_values: Dict[str, str],
    \newline
    total\_count: int,
    \newline
    unique\_types: int,
    \newline
    message: str,
    \newline
    error: str \\
    \hline

    var\_get\_multiple & Get values of multiple simulation variables at once. &
    variable\_paths: List[str] & success: bool,
    \newline
    variables: Dict[str, Any],
    \newline
    set\_variables: int,
    \newline
    errors: List[str],
    \newline
    simulation: str \\
    \hline

    var\_set\_multiple & Set values of multiple simulation variables at once. &
    variable\_paths: List[str] & success: bool,
    \newline
    variables: Dict[str, Any],
    \newline
    set\_variables: int,
    \newline
    errors: List[str],
    \newline
    simulation: str \\

    param\_set\_multiple & Set values of multiple simulation parameters at once.
    & parameter\_paths: List[str] & success: bool,
    \newline
    simulation: str,
    \newline
    updated\_parameters: int,
    \newline
    total\_requested: int,
    \newline
    results: Dict[str, Any],
    \newline
    errors: str,
    \newline
    message: str \\
    \hline

    model\_add & Add a model to the APS simulation. & model\_type: str
    \newline
    x: float, \newline y: float & success: bool,
    \newline
    model\_name: str,
    \newline
    model\_type: str,
    \newline
    full\_model\_type: str,
    \newline
    position: Dict[str, float],
    \newline
    message: str,
    \newline
    error: str \\

    models\_connect & Creates a connection between two model ports and validates
    that the connection is real. & start\_port: str, end\_port: str & success:
    bool,
    \newline
    connection: str,
    \newline
    connector\_name: str,
    \newline
    error: str,
    \newline
    warning: str \\
    \hline

    fluid\_create & Create a new fluid model in APS with the desired components and
    thermodynamic settings. & library\_name: str,
    \newline
    fluid\_name: str,
    \newline
    components: List[str], thermo\_method: str,
    \newline
    phases: str & success: bool,
    \newline
    library\_name: str,
    \newline
    fluid\_width\_name: str,
    \newline
    components\_requested: List[str],
    \newline
    components\_added: int,
    \newline
    thermo\_method: str,
    \newline
    phases: str,
    \newline
    fluid\_existed: bool,
    \newline
    results: List[str],
    \newline
    warnings: List[str],
    \newline
    message: str,
    \newline
    error: str \\

    fluid\_to\_source & Set fluid type of one specific source model. & fluid\_name:
    str, source\_name: str & success: bool,
    \newline
    simulation: str,
    \newline
    source\_name: str,
    \newline
    fluid\_name: str,
    \newline
    fluid\_type\_path: str,
    \newline
    parameter\_path: str,
    \newline
    message: str,
    \newline
    error: str \\
\end{longtable}

\section{Prompts}
\label{app:prompts}

\subsection{Case Study 1: Analysis of the Simulation Flowsheet}

\begin{framed}
    \textbf{Prompt 1.0:}

    \textit{\textbf{Role}: You are a process simulation expert specializing in
    chemical process modeling using AVEVA Process Simulation (APS). You can use the
    predefined APS toolset to connect, open, modify, inspect, and analyze simulations.}\\
    \textit{\textbf{Task}: Analyze the flowsheet named “C1 - Water Methanol
    Separation”}\\
    \textit{\textbf{Required Steps}:\\
    - Connect to APS.\\
    - Open the specified simulation file.\\
    - Retrieve the flowsheet topology, including all models and connectors.\\
    - Read and report key variables and parameters from relevant models (equipment,
    streams, controllers, etc.).\\
    - Produce a consolidated summary of all extracted information.} \\
    \textit{\textbf{Return Format}: Provide a short bullet-point summary (maximum
    500 words)\\
    - Major equipment and stream connectivity \\
    - Important operating conditions or parameters \\
    - Any noteworthy observations from the simulation data}
\end{framed}

\begin{framed}
    \textbf{Answer 1.0:}

    \textit{I'll analyze the "C1 - Water Methanol Separation" flowsheet for you.
    Let me start by connecting to APS and opening the simulation.}\\
    \textit{$\longrightarrow$ \textbf{aps\_connect}()\textrightarrow(connection: true, ...)} \\
    \textit{Great! APS is connected and the simulation is already open. Now let
    me retrieve the flowsheet topology to understand the models and connectors.}\\
    \textit{$\longrightarrow$ \textbf{models\_list}()\textrightarrow(success: true, ...)} \\
    \textit{$\longrightarrow$ \textbf{connectors\_list}()\textrightarrow(success: true, ...)} \\
    $\longrightarrow$ \textit{\textbf{model\_all\_vars}(model\_name: T1)\textrightarrow(success: true, ...)}\\
    $\longrightarrow$ \textit{\textbf{model\_all\_vars}(model\_name: MeOH\_Water)\textrightarrow(success: true, ...)}\\
    \textit{Now let me get the key stream variables to understand the flows and
    compositions:}\\
    $\longrightarrow$ \textit{\textbf{var\_get\_multiple}(variables: [S1.F, S1.W,
    S1.T, S1.P, S1.z[METHANOL], S1.z[WATER], S2.F, S2.W, S2.T, \newline \hspace*{.6cm} S2.P,
    S2.z[METHANOL], S2.z[WATER], S3.F, S3.W, S3.T, S3.P, S3.z[METHANOL], S3.z[WATER]])\textrightarrow \newline \hspace*{.6cm}(success: true, ...)}\\
    \textit{Let me also check the simulation status to see if it's converged:}\\
    \textit{$\longrightarrow$ \textbf{sim\_status}()\textrightarrow(success: true, ...)}

    \textit{\textbf{Analysis Summary: C1 - Water Methanol Separation Flowsheet}}\\
    \textit{\textbf{Major Equipment and Stream Connectivity:}\\
    - Column T1: 13-stage distillation column (2.03 m diameter, 7.67 m total height)
    with internal condenser and reboiler\\
    - Feed Source: MeOH\_Water source feeding stage 7 of the column\\
    - Product Sinks: DIST (distillate) and BOTTOMS (bottoms) collection points\\
    - Stream Configuration: S3 (feed) → T1 → S1 (distillate) and S2 (bottoms)\\
    \textbf{Important Operating Conditions:}\\
    - Feed Stream (S3): 8.79 kg/s (0.397 kmol/s) at 351.15 K and 150 kPa, containing
    29.4 mol\% methanol and 70.6 mol\% water\\
    - Distillate (S1): 4.13 kg/s (0.139 kmol/s) at 339.8 K, highly enriched in
    methanol (84.2 mol\% methanol, 15.8 mol\% water)\\
    - Bottoms (S2): 4.66 kg/s (0.259 kmol/s) at 379.9 K, essentially pure water
    (99.998 mol\% water, trace methanol)\\
    - Column Pressure: Top at 101.3 kPa, bottom at 127.3 kPa (2 kPa/stage
    pressure drop)\\
    \textbf{Key Performance Data:}\\
    - Condenser Duty: -10,091 kJ/s (cooling requirement)\\
    - Reboiler Duty: +10,500 kJ/s (heating requirement)\\
    - Reflux Ratio: 1.0 (equal reflux to distillate)\\
    - Temperature Profile: 344-379 K from top to bottom stages\\
    - Separation Efficiency: Excellent methanol recovery in distillate with high
    purity water bottoms\\
    \textbf{Noteworthy Observations:}\\
    - The simulation is fully converged and properly specified\\
    - Feed enters at stage 7 (middle of 13-stage column), optimal for this
    separation\\
    - Very low methanol concentration in bottoms (0.002 mol\%) indicates
    excellent separation\\
    - Energy balance is satisfied (slight difference between condenser and
    reboiler duties accounts for sensible heat effects)\\
    - Standard tray efficiency of 100\% assumed throughout the column}
\end{framed}

\begin{framed}
    \textbf{Prompt 1.1:}

    \textit{What options are there to improve the separation?}
\end{framed}

\begin{framed}
    \textbf{Answer 1.1:}

    \textit{I'll analyze the current separation performance and identify
    improvement options. Let me first get some additional key parameters to better
    understand the optimization opportunities.}\\
    $\longrightarrow$ \textit{\textbf{model\_all\_params}(model\_name: T1)\textrightarrow(success: true, ...)}\\
    \textit{Based on my analysis of the current methanol-water separation system,
    here are the main options to improve the separation:}

    \textit{\textbf{Process Configuration Improvements}\\
    \textbf{1. Increase Number of Stages}\\
    - Current: 13 theoretical stages\\
    - Option: Add 2-5 more stages (15-18 total) to improve separation efficiency\\
    - Benefit: Higher purity products with potentially lower reflux ratio\\
    \textbf{2. Optimize Feed Stage Location}\\
    - Current: Stage 7 (middle of column) - Option: Test feed stages 6-8 to find
    optimal thermal efficiency\\
    - Benefit: Minimize reboiler duty while maintaining product quality\\
    \textbf{3. Add Side Streams}\\
    - Current: Simple distillation with overhead/bottoms\\
    - Option: Add side draws for intermediate compositions\\
    - Benefit: Better heat integration and product flexibility}

    \textit{\textbf{Operating Parameter Optimization}\\
    \textbf{4. Reflux Ratio Adjustment}\\
    - Current: RR = 1.0 (conservative)\\
    - Option: Reduce to 0.7-0.8 or increase to 1.2-1.5 depending on objectives\\
    - Lower RR: Reduces energy consumption but may compromise purity\\
    - Higher RR: Improves separation but increases energy costs\\
    \textbf{5. Pressure Optimization}\\
    - Current: 101.3 kPa top pressure\\
    - Option: Operate at higher pressure (150-200 kPa) for better heat integration\\
    - Benefit: Higher condensing temperature enables heat recovery\\
    \textbf{6. Feed Conditioning}\\
    - Current: Feed at 351K (subcooled liquid, VF = -0.33)\\
    - Option: Preheat feed closer to bubble point or partial vaporization\\
    - Benefit: Reduces reboiler duty requirements}

    \textit{\textbf{Advanced Process Configurations}\\
    \textbf{7. Heat Integration}\\
    - Install feed-effluent heat exchangers\\
    - Use hot bottoms product to preheat feed\\
    - Benefit: 15-25\% energy savings typically achievable\\
    \textbf{8. Double-Effect Configuration}\\
    - Operate two columns at different pressures\\
    - Use overhead vapor from high-pressure column to drive low-pressure
    reboiler\\
    - Benefit: \~40\% energy reduction possible\\
    \textbf{9. Reactive Distillation}\\
    - If methanol purification is the goal, consider reactive separation\\
    - Benefit: Simultaneous reaction and separation}

    \textit{\textbf{Equipment Modifications}\\
    \textbf{10. Enhanced Tray Design}\\
    - Current: 100\% efficiency assumption\\
    - Option: Install high-efficiency trays (structured packing)\\
    - Benefit: Better mass transfer, potentially fewer actual trays needed\\
    \textbf{11. Advanced Control Systems}\\
    - Implement composition control on key streams\\
    - Add feedforward control for feed disturbances\\
    - Benefit: More consistent product quality and energy efficiency}

    \textit{\textbf{Immediate Quick Wins}\\
    \textbf{Most promising near-term improvements:}\\
    - Optimize reflux ratio (test 0.8-1.2 range)\\
    - Preheat feed to reduce subcooling\\
    - Add 2-3 stages if column modifications are feasible\\
    - Install basic heat integration between hot and cold streams\\
    \textbf{Energy vs. Separation Trade-offs:}\\
    - Current energy consumption: ~20.6 MJ/h total (condenser + reboiler)\\
    - Potential energy savings: 20-40\% with heat integration\\
    - Purity improvements: Can achieve >95\% methanol and >99.99\% water with optimization}

    \textit{Would you like me to simulate any specific optimization scenarios or
    analyze the economics of these improvements?}
\end{framed}

\begin{framed}
    \textbf{Prompt 1.2:}

    \textit{Try adjusting the reflux ratio to achieve a methanol purity above 95
    mol\% in the distillate stream using the current column configuration.}
\end{framed}

\begin{framed}
    \textbf{Answer 1.2:} \raggedright

    \textit{I'll adjust the reflux ratio to achieve >95 mol\% methanol purity in
    the distillate. Let me first check the current reflux ratio parameter and then
    systematically increase it.}\\
    $\longrightarrow$ \textit{\textbf{var\_get\_multiple}(variables: [T1.RefluxRatio`,
    T1.ReboilRatio, S1.z[METHANOL]])\textrightarrow(success: true, ...)}\\
    \textit{Current methanol purity is 84.2 mol\%, so I need to increase the
    reflux ratio. Let me try increasing it to 1.5 first:}\\
    $\longrightarrow$ \textit{\textbf{var\_set\_multiple}(variable\_data: [{path: T1.RefluxRatio, value: 1.5}])\textrightarrow(success: true, ...)}\\
    \textit{$\longrightarrow$ \textbf{sim\_status}}()\textrightarrow(...)\\
    $\longrightarrow$ \textit{\textbf{var\_get\_multiple}(variables: [S1.z[METHANOL],
    S1.z[WATER], S1.F, S1.W, T1.DutyCond, T1.DutyReb])\textrightarrow \newline \hspace*{.6cm}(success: true, ...)}\\
    \textit{Excellent! With a reflux ratio of 1.5, I achieved 96.3 mol\% methanol
    purity, which exceeds the target of 95 mol\%. Let me also check the bottoms
    composition to ensure we haven't compromised the water purity:}\\
    $\longrightarrow$ \textit{\textbf{var\_get\_multiple}(variables: [S2.z[METHANOL],
    S2.z[WATER], S2.F, S2.W])\textrightarrow(success: true, ...)}\\
    \textit{Perfect! Let me also check if we can achieve exactly 95\% with a
    slightly lower reflux ratio to minimize energy consumption:}\\
    $\longrightarrow$ \textit{\textbf{var\_set\_multiple}(variable\_data: [{path: T1.RefluxRatio, value: 1.3}])\textrightarrow(success: true, ...)}\\
    $\longrightarrow$ \textit{\textbf{var\_get\_multiple}(variables: [S1.z[METHANOL],
    T1.DutyCond, T1.DutyReb])\textrightarrow(success: true, ...)}\\
    \textit{That gives 91.5\% methanol, which is below target. Let me try 1.4:}\\
    $\longrightarrow$ \textit{\textbf{var\_set\_multiple}(variable\_data: [{path: T1.RefluxRatio, value: 1.4}])\textrightarrow(success: true, ...)}\\
    $\longrightarrow$ \textit{\textbf{var\_get\_multiple} $\longrightarrow$ ERROR: Tool
    'havia:var\_get\_multiple' not found.}\\
    $\longrightarrow$ \textit{\textbf{var\_get\_multiple}(variables: [S1.z[METHANOL], S1.z[WATER], S1.F, S1.W, S2.z[METHANOL], \newline \hspace*{.6cm} S2.z[WATER], T1.DutyCond, T1.DutyReb])\textrightarrow(success: true, ...)}\\
    \textit{Close! That gives 93.9\% methanol. Let me try 1.45 to get closer to 95\%:}\\
    $\longrightarrow$ \textit{\textbf{var\_set\_multiple}(variable\_data: [{path: T1.RefluxRatio, value: 1.45}])\textrightarrow(success: true, ...)}\\
    $\longrightarrow$ \textit{\textbf{var\_get\_multiple}(variables: [S1.z[WATER],
    S1.F, S1.W, S2.z[METHANOL], S2.z[WATER], T1.DutyCond, \hspace*{.6cm} T1.DutyReb,
    T1.RefluxRatio])\textrightarrow(success: true, ...)}\\
    \textit{Perfect! With a reflux ratio of 1.45, I've achieved 95.1 mol\% methanol
    purity in the distillate, which meets the target of >95 mol\%.}

    \textit{\textbf{Optimization Results Summary}\\
    Successfully achieved >95 mol\% methanol purity with reflux ratio = 1.45}

    \textit{\textbf{Performance Comparison:}}\\
    \begin{tabular}{llll}
        \hline
        \multicolumn{1}{l}{\textit{\textbf{Parameter}}} & \multicolumn{1}{l}{\textit{\textbf{Original (RR=1.0)}}} & \multicolumn{1}{l}{\textit{\textbf{Optimized (RR=1.45)}}} & \multicolumn{1}{l}{\textit{\textbf{Change}}} \\
        \hline
        \textit{\textbf{Distillate Methanol Purity}}    & \textit{84.2 mol\%}                                     & \textit{\textbf{95.1 mol\%}}                              & \textit{\textbf{+10.9\%}}                    \\
        \textit{\textbf{Distillate Flow Rate}}          & \textit{0.139 kmol/s}                                   & \textit{0.123 kmol/s}                                     & \textit{-11.2\%}                             \\
        \textit{\textbf{Distillate Mass Flow}}          & \textit{4.13 kg/s}                                      & \textit{3.85 kg/s}                                        & \textit{-6.8\%}                              \\
        \textit{\textbf{Bottoms Water Purity}}          & \textit{99.998 mol\%}                                   & \textit{99.998 mol\%}                                     & \textit{No change}                           \\
        \textit{\textbf{Condenser Duty}}                & \textit{-10,091 kJ/s}                                   & \textit{-10,711 kJ/s}                                     & \textit{+6.1\%}                              \\
        \textit{\textbf{Reboiler Duty}}                 & \textit{+10,500 kJ/s}                                   & \textit{+11,145 kJ/s}                                     & \textit{+6.1\%}                              \\
        \hline
    \end{tabular}

    \textit{\textbf{Key Observations:}\\
    - \textbf{Target Achieved:} Methanol purity increased from 84.2\% to 95.1
    mol\%\\
    - \textbf{Water Quality Maintained:} Bottoms water purity remains excellent at
    99.998 mol\%\\
    - \textbf{Energy Trade-off:} Energy consumption increased by ~6\% (additional
    645 kJ/s)\\
    - \textbf{Yield Impact:} Slightly lower distillate flow rate due to higher
    selectivity\\
    \textbf{Economic Impact:}\\
    - \textbf{Product Value:} Higher purity methanol typically commands 10-15\% price
    premium\\
    - \textbf{Energy Cost:} 6\% increase in utilities cost\\
    - \textbf{Net Benefit:} Positive due to premium pricing for high-purity
    methanol}

    \textit{The optimization successfully demonstrates that the current 13-stage
    column can achieve >95\% methanol purity with a moderate increase in reflux
    ratio, making this an excellent operational improvement option.}
\end{framed}

\subsection{Case Study 2: Synthesis of the Simulation Flowsheet}
\label{app:case2}

\begin{framed}
    \textbf{Prompt 2.0.0:}

    \textit{\textbf{Role}: You are a process simulation expert specializing in
    chemical process modeling using AVEVA Process Simulation (APS). You can use the
    predefined APS toolset to connect, open, modify, inspect, and analyze simulations.\\
    \textbf{Task}: Follow exactly my steps to construct a simulation model for the
    separation of water and methanol\\
    \textbf{First Steps}: \\
    1. Connect to APS \\
    2. Create simulation "Sep\_MeOHWater" \\
    3. Open simulation "Sep\_MeOHWater" \\}
\end{framed}

\begin{framed}
    \textbf{Answer 2.0.0:}

    \textit{I'll help you construct a simulation model for the separation of
    water and methanol using AVEVA Process Simulation. Let me follow your steps exactly.}\\
    \textit{\textbf{Step 1: Connect to APS}}\\
    \textit{$\longrightarrow$ \textbf{aps\_connect}()\textrightarrow(connection: true, ...)} \\
    \textit{Great! Successfully connected to APS. I can see there's currently
    one open simulation.}\\
    \textit{\textbf{Step 2: Create simulation "Sep\_MeOHWater"}}\\
    $\longrightarrow$ \textit{\textbf{sim\_create}(sim\_name: Sep\_MeOHWater)\textrightarrow(success: true, ...)}
    \\
    \textit{Excellent! Successfully created the new simulation "Sep\_MeOHWater".}\\
    \textit{\textbf{Step 3: Open simulation "Sep\_MeOHWater"}}\\
    $\longrightarrow$ \textit{\textbf{sim\_open}(sim\_name: Sep\_MeOHWater)\textrightarrow(success: true, ...)}
    \\
    \textit{Perfect! All three initial steps completed successfully:\\
    \checkmark Step 1: Connected to APS\\
    \checkmark Step 2: Created simulation "Sep\_MeOHWater"\\
    \checkmark Step 3: Opened simulation "Sep\_MeOHWater"}\\
    \textit{The simulation "Sep\_MeOHWater" is now ready for further model
    construction. What would you like me to do next for the water-methanol separation
    process?}
\end{framed}

\begin{framed}
    \textbf{Prompt 2.0.1:}

    \textit{\textbf{Next Step}: Create fluid with name "MeOHWater" with
    components "metahnol" and "water", method "NRTL", phases "VLE".}
\end{framed}

\begin{framed}
    \textbf{Answer 2.0.1:}

    \textit{\textbf{Next Step: Create fluid with name "MeOHWater"}}\\
    $\longrightarrow$ \textit{\textbf{fluid\_create}(phases: Vapor/Liquid (VLE),
    components: [methanol, water], fluid\_name: MeOHWater,
    \newline
    \hspace*{0.6cm} library\_name: Sep\_MeOHWater, thermo\_method: Non-Random Two-Liquid
    (NRTL))\textrightarrow(success: true, ...)} \\
    \textit{\checkmark \textbf{Fluid Creation Completed Successfully!}}\\
    \textit{The fluid package "MeOHWater" has been created with:\\
    - \textbf{Components}: methanol and water (2 components added successfully)\\
    - \textbf{Thermodynamic Method}: Non-Random Two-Liquid (NRTL)\\
    - \textbf{Phases}: Vapor/Liquid (VLE)\\
    - \textbf{Library}: Sep\_MeOHWater}\\
    \textit{The fluid package is now ready for use in the simulation. What's the
    next step for building the water-methanol separation process?}
\end{framed}

\begin{framed}
    \textbf{Prompt 2.0.2:}

    \textit{\textbf{Next Step}: Add a source model to the flowsheet.}
\end{framed}

\begin{framed}
    \textbf{Answer 2.0.2:}

    \textit{\textbf{Next Step: Add source model to the flowsheet}}\\
    $\longrightarrow$ \textit{\textbf{model\_add}(model\_type: Source)\textrightarrow(success: true, ...)}
    \\
    \textit{\checkmark \textbf{Source Model Added Successfully!}}\\
    \textit{The source model has been added to the flowsheet with:\\
    - \textbf{Model Name}: SRC1\\
    - \textbf{Model Type}: Source\\
    - \textbf{Position}: (100, 100) on the flowsheet}\\
    \textit{The source model is now available on the flowsheet and ready for
    configuration. What's the next step for the water-methanol separation process?}
\end{framed}

\begin{framed}
    \textbf{Prompt 2.0.3:}

    \textit{\textbf{Next Step}: Set fluid of source to "MeOHWater".}
\end{framed}

\begin{framed}
    \textbf{Answer 2.0.3:}

    \textit{\textbf{Set fluid of source to "MeOHWater"}}\\
    $\longrightarrow$ \textit{\textbf{fluid\_to\_source}(fluid\_name: MeOHWater,
    source\_name: SRC1)\textrightarrow(success: true, ...)} \\
    \textit{\checkmark \textbf{Fluid Assignment Completed Successfully!}}\\
    \textit{The fluid package "MeOHWater" has been assigned to source model "SRC1".
    The source is now configured to use the methanol-water fluid package with NRTL
    thermodynamics and VLE phases.}\\
    \textit{What's the next step for building the separation process?}
\end{framed}

\begin{framed}
    \textbf{Prompt 2.0.4:}

    \textit{\textbf{Next Step}: Add a distillation column to the fl owsheet, and
    connect the source to the column, port name "Fin"}
\end{framed}

\begin{framed}
    \textbf{Answer 2.0.4:}

    \textit{\textbf{Add distillation column to the flowsheet}}\\
    $\longrightarrow$ \textit{\textbf{model\_add}(x: 300, y: 100, model\_type:
    Column)\textrightarrow(success: true, ...)} \\
    \textit{\checkmark \textbf{Column Added Successfully!}}\\
    \textit{Now connecting the source to the column at port "Fin":}\\
    $\longrightarrow$ \textit{\textbf{models\_connect}(to\_port: T1.Fin, from\_port:
    SRC1.Out)\textrightarrow(success: true, ...)} \\
    \textit{\checkmark \textbf{Connection Completed Successfully!}}\\
    \textit{- \textbf{Distillation Column}: T1 added at position (300, 100)\\
    - \textbf{Connection}: SRC1.Out → T1.Fin\\
    - \textbf{Stream Name}: S1\\
    - \textbf{Current Flow}: 0.031208882297572946}\\
    \textit{The source is now connected to the distillation column's feed port. What's
    the next step for configuring the separation process?}
\end{framed}

\begin{framed}
    \textbf{Prompt 2.0.5:}

    \textit{\textbf{Next Step}: Set the parameters of columns: NStages: 13, FeedStage:
    7, Condenser: Internal, Reboiler: Internal, ReferenceStreams: Typical, StaticHead:
    Include}
\end{framed}

\begin{framed}
    \textbf{Answer 2.0.5:}

    \textit{\textbf{Set the column parameters}}\\
    $\longrightarrow$ \textit{\textbf{param\_set\_multiple}(parameter\_data: [{path: T1.NStages, value: 13},
    {path: T1.FeedStage[S1], value: 7}, {path: \newline \hspace*{.6cm} T1.Condenser, value: Internal},
    {path: T1.Reboiler, value: Internal}, {path: T1.ReferenceStreams, value: Typical},
    \newline
    \hspace*{.6cm} {path: T1.StaticHead, value: Include}])\textrightarrow(success: true, ...)}

    \textit{\checkmark \textbf{Column Parameters Set Successfully!}}\\
    \textit{All column parameters have been configured:\\
    - \textbf{NStages}: 13 stages\\
    - \textbf{FeedStage}: Stage 7 (for stream S1)\\
    - \textbf{Condenser}: Internal\\
    - \textbf{Reboiler}: Internal\\
    - \textbf{ReferenceStreams}: Typical\\
    - \textbf{StaticHead}: Include}\\
    \textit{The distillation column T1 is now properly configured with 13 stages,
    feed entering at stage 7, and internal condenser and reboiler. What's the next
    step for completing the separation process?}
\end{framed}

\begin{framed}
    \textbf{Prompt 2.0.6:}

    \textit{\textbf{Next Step}: Add model sink with name "Distillate" and
    connect from the column port name "Dout". Add model sink with name "Bottoms"
    and connect from the column port name "Lout".}
\end{framed}

\begin{framed}
    \textbf{Answer 2.0.6:}

    \textit{\textbf{Add sink models "Distillate" and connect to column port "Dout"}}\\
    $\longrightarrow$ \textit{\textbf{model\_add}(x: 400, y: 50, model\_name:
    Distillate, model\_type: Sink)\textrightarrow(success: true, ...)} \\
    $\longrightarrow$ \textit{\textbf{models\_connect}(to\_port: Distillate.In, from\_port:
    T1.Dout)\textrightarrow(success: true, ...)} \\
    \textit{\textbf{Add sink model "Bottoms" and connect from column port "Lout"}}\\
    $\longrightarrow$ \textit{\textbf{model\_add}(x: 400, y: 150, model\_name:
    Bottoms, model\_type: Sink)\textrightarrow(success: true, ...)} \\
    $\longrightarrow$ \textit{\textbf{models\_connect}(to\_port: Bottoms.In, from\_port:
    T1.Lout)\textrightarrow(success: true, ...)} \\
    \textit{\checkmark \textbf{Both Sink Models Added and Connected Successfully!}}\\
    \textit{\textbf{Distillate Sink:}\\
    - \textbf{Model Name}: Distillate\\
    - \textbf{Position}: (400, 50)\\
    - \textbf{Connection}: T1.Dout → Distillate.In (Stream S2, Flow: 0.0105)}\\
    \textit{\textbf{Bottoms Sink:}\\
    - \textbf{Model Name}: Bottoms\\
    - \textbf{Position}: (400, 150)\\
    - \textbf{Connection}: T1.Lout → Bottoms.In (Stream S3, Flow: 0.0207)}\\
    \textit{The distillation column is now fully connected with feed input and
    both distillate and bottoms product streams. What's the next step for completing
    the water-methanol separation simulation?}
\end{framed}

\begin{framed}
    \textbf{Prompt 2.0.7:}

    \textit{\textbf{Next Step}: Set variables in the source model:\\
    - set variable Source.W to 10 kg/s\\
    - set variable Source.T to 80 °C\\
    - set variable Source.P to 1 bar\\
    - set variable Source.M[methanol] to 0.5\\
    - set variable Source.M[water] to 0.5\\
    - set variable Column.RefluxRatio to 1.5\\
    - set variable Column.BoilupRatio to 1.2}
\end{framed}

\begin{framed}
    \textbf{Answer 2.0.7:}

    \textit{\textbf{Set variables in the source model and column}}\\
    $\longrightarrow$ \textit{\textbf{var\_set\_multiple}(variable\_data: [{path: SRC1.W, unit: kg/s, value: 10},
    {path: SRC1.T, unit: C, value: 80},
    \newline
    \hspace*{0.6cm} {path: SRC1.P, unit: bar, value: 1},
    {path: SRC1.M[methanol], value: 0.5}, {path: SRC1.M[water], value: 0.5},
    \newline
    \hspace*{0.6cm} {path: T1.RefluxRatio, value: 1.5}, {path: T1.BoilupRatio, value: 1.2}])\textrightarrow(success: true, ...)}\\
    \textit{\checkmark \textbf{All Variables Set Successfully!}}\\
    \textit{\textbf{Source Model (SRC1) Variables:}\\
    - \textbf{W}: 10 kg/s (mass flow rate)\\
    - \textbf{T}: 80 °C (temperature)\\
    - \textbf{P}: 1 bar (pressure)\\
    - \textbf{M[methanol]}: 0.5 (mole fraction)\\
    - \textbf{M[water]}: 0.5 (mole fraction)}\\
    \textit{\textbf{Column Model (T1) Variables:}\\
    - \textbf{RefluxRatio}: 1.5\\
    - \textbf{BoilupRatio}: 1.2}\\
    \textit{The source is now configured with a 50/50 mole fraction methanol-water
    mixture at 80°C, 1 bar, and 10 kg/s flow rate. The distillation column is set
    with reflux ratio of 1.5 and boilup ratio of 1.2.}\\
    \textit{What's the next step? Would you like me to run the simulation or make
    any additional configurations?}
\end{framed}

\begin{framed}
    \textbf{Prompt 2.0.8:}

    \textit{Now please inspect the key variables in the models (source, column, sink)
    and give me a brief summary.}
\end{framed}

\begin{framed}
    \textbf{Answer 2.0.8:}

    \textit{\textbf{Inspecting key variables in all models}}\\
    \textit{Let me check the key variables for each model in the simulation.}\\
    $\longrightarrow$ \textit{\textbf{model\_all\_vars}(model\_name: SRC1)\textrightarrow(success: true, ...)}\\
    $\longrightarrow$ \textit{\textbf{model\_all\_vars}(model\_name: T1)\textrightarrow(success: true, ...)}\\
    $\longrightarrow$ \textit{\textbf{model\_all\_vars}(model\_name: Distillate)\textrightarrow(success: true, ...)}\\
    $\longrightarrow$ \textit{\textbf{model\_all\_vars}(model\_name: Bottoms)\textrightarrow(success: true, ...)}

    \textit{\textbf{Key Variables Summary}}\\
    \textit{\textbf{Source Model (SRC1)}\\
    - \textbf{Feed Conditions}: 10.0 kg/s, 80°C (353.15 K), 1 bar (100 kPa)\\
    - \textbf{Composition}: 50/50 mole\% methanol/water feed\\
    - \textbf{Flow Properties}: 0.4 kmol/s molar flow, partially vaporized (VF =
    0.69)\\
    - \textbf{Mass Fractions}: 64\% methanol, 36\% water by mass}\\
    \textit{\textbf{Distillation Column (T1)}\\
    - \textbf{Operating Conditions}: 13 stages, feed at stage 7, atmospheric
    pressure\\
    - \textbf{Temperature Profile}: Top = 67°C (340 K), Bottom = 100°C (373 K)\\
    - \textbf{Separation Performance}: Good separation with temperature gradient\\
    - \textbf{Duties}: Condenser = -19,309 kJ/s, Reboiler = 9,010 kJ/s\\
    - \textbf{Ratios}: Reflux = 1.5, Boilup = 1.2 (as specified)}\\
    \textit{\textbf{Distillate Product (Distillate Sink)}\\
    - \textbf{Product Rate}: 6.7 kg/s, 0.22 kmol/s\\
    - \textbf{Temperature}: 65°C (338 K), subcooled liquid (VF = -0.07)\\
    - \textbf{Composition}: \textbf{92.6\% methanol}, 7.4\% water (excellent purity!)\\
    - \textbf{Recovery}: High methanol recovery in overhead product}\\
    \textit{\textbf{Bottoms Product (Bottoms Sink)}\\
    - \textbf{Product Rate}: 3.3 kg/s, 0.18 kmol/s\\
    - \textbf{Temperature}: 100°C (373 K), essentially at boiling point\\
    - \textbf{Composition}: \textbf{99.999\% water}, trace methanol (excellent purity!)\\
    - \textbf{Recovery}: Nearly complete water recovery in bottoms}

    \textit{\textbf{Overall Performance}\\
    The separation is working excellently with high-purity products:\\
    - \textbf{Distillate}: 92.6\% methanol (light component concentrated
    overhead)\\
    - \textbf{Bottoms}: 99.999\% water (heavy component concentrated at bottom)\\
    - \textbf{Good mass balance}: 10 kg/s in = 6.7 + 3.3 kg/s out}\\
    \textit{The column is operating efficiently for the methanol-water
    separation with the specified operating conditions.}
\end{framed}

\begin{framed}
    \textbf{Prompt 2.1.0:}

    \textit{Construct a simulation model for the separation of water (50 mol\%) and
    methanol (50 mol\%) for a feed at at 80 °C, 1 bar with a fl owrate of 10 kg/s
    using AVEVA Process Simulation.\\
    - Use the tool to get the correct variable and parameter names before setting,
    to prevent modifying the wrong variable names.\\
    - Ask the user to connect the models, do not connect yourself.\\
    - Quit trying if you meet troubles more than twice.}
\end{framed}

\begin{framed}
    \textbf{Answer 2.1.0:}

    \textit{I'll help you construct a water-methanol separation simulation in
    AVEVA Process Simulation. Let me start by connecting to APS and creating the
    simulation.}\\
    \textit{$\longrightarrow$ \textbf{aps\_connect}()\textrightarrow(connection: true, ...)} \\
    \textit{Good! I'm connected to APS. I can see several water-methanol
    separation simulations already exist. Let me create a new simulation for your
    specifi c requirements.}\\
    $\longrightarrow$ \textit{\textbf{sim\_create}(sim\_name: WaterMethanol\_Sep\_80C\_10kgs\_v3)\textrightarrow(success: true, ...)}
    \\
    \textit{Perfect! Now let me create the fl uid package for water and methanol
    with NRTL thermodynamics:}\\
    $\longrightarrow$ \textit{\textbf{fluid\_create}(phases: Vapor/Liquid (VLE),
    components: [Water, Methanol], fluid\_name:
    \newline
    \hspace*{0.6cm} WaterMethanol\_NRTL, library\_name: WaterMethanol\_Sep\_80C\_10kgs\_v3,
    thermo\_method: Non-Random
    \newline
    \hspace*{0.6cm} Two-Liquid (NRTL))\textrightarrow(success: true, ...)} \\
    \textit{Excellent! Now I'll add the necessary models for the separation fl owsheet:}\\
    \textbf{1. Feed Source} - for the feed stream\\
    \textbf{2. Distillation Column} - for separation\\
    \textbf{3. Two Sinks} - for distillate and bottoms products\\
    $\longrightarrow$ \textit{\textbf{model\_add}(x: 100, y: 200, model\_name:
    Feed, model\_type: Source)\textrightarrow(success: true, ...)}\\
    $\longrightarrow$ \textit{\textbf{model\_add}(x: 300, y: 200, model\_name:
    Column, model\_type: Column)\textrightarrow(success: true, ...)}\\
    $\longrightarrow$ \textit{\textbf{model\_add}(x: 450, y: 150, model\_name:
    Distillate, model\_type: Sink)\textrightarrow(success: true, ...)}\\
    $\longrightarrow$ \textit{\textbf{model\_add}(x: 450, y: 250, model\_name:
    Bottoms, model\_type: Sink)\textrightarrow(success: true, ...)}\\
    \textit{Now let me assign the fluid package to the feed source:}\\
    $\longrightarrow$ \textit{\textbf{fluid\_to\_source}(fluid\_name:
    WaterMethanol\_NRTL, source\_name: Feed)\textrightarrow(success: true, ...)} \\
    \textit{Great! Now I need to get the correct variable names before setting the
    feed conditions. Let me check the Feed model variables:}\\
    $\longrightarrow$ \textit{\textbf{model\_all\_vars}(model\_name: Feed)\textrightarrow(success: true, ...)}\\
    \textit{Perfect! Now I can see the variable names. Let me set the feed
    conditions:\\
    - Mass flow rate: W = 10 kg/s\\
    - Temperature: T = 353.15 K (80°C)\\
    - Pressure: P = 101.325 kPa (1 bar)\\
    - Composition: 50 mol\% each (using M[Water] = M[Methanol] = 1.0, Mt = 2.0)}\\
    $\longrightarrow$ \textit{\textbf{var\_set\_multiple}(variable\_data: [{path: Feed.W`,unit: kg/s, value: 10},{ path: Feed.T, unit: K, value: \newline \hspace*{0.6cm} 353.15},
    {path: Feed.P, unit: kPa, value: 101.325},{path: Feed.M[Water], value: 1},{path: Feed.M[Methanol], \newline \hspace*{0.6cm} value: 1},{path: Feed.Mt, value: 2}])\textrightarrow(success: true, ...)}\\
    \textit{Excellent! Now let me check the Column model parameters to configure
    it properly:}\\
    $\longrightarrow$ \textit{\textbf{model\_all\_params}(model\_name: Column)\textrightarrow(success: true, ...)}\\
    \textit{Now let me configure the column with basic parameters - I'll set it
    up with condenser, reboiler, and appropriate number of stages:}\\
    $\longrightarrow$ \textit{\textbf{param\_set\_multiple}(parameter\_data: [{path: Column.FluidType, value: \newline \hspace*{0.6cm} WaterMethanol\_Sep\_80C\_10kgs\_v3.Models.WaterMethanol\_NRTL},
    { path: Column.NStages, value: 15}, {path: \newline \hspace*{0.6cm} Column.Condenser, value: Internal},
    {path: Column.Reboiler, value: Internal}])\textrightarrow(success: true, ...)}

    \textit{Great! Now I have set up the basic flowsheet with:}\\
    \textit{\textbf{1. Feed Source} (Feed) - configured with:\\
    - 10 kg/s mass flow rate\\
    - 80°C (353.15 K) temperature\\
    - 1 bar (101.325 kPa) pressure\\
    - 50 mol\% Water, 50 mol\% Methanol\\
    - WaterMethanol\_NRTL fluid package}\\
    \textit{\textbf{2. Distillation Column} (Column) - configured with:\\
    - 15 theoretical stages\\
    - Internal condenser\\
    - Internal reboiler\\
    - WaterMethanol\_NRTL fluid package}\\
    \textit{\textbf{3. Product Sinks} - Distillate and Bottoms}

    \textit{\textbf{Now I need you to connect the models manually using the
    AVEVA interface:}}\\
    \textit{\textbf{1. Connect the Feed to Column:}\\
    - Connect \texttt{Feed.Out} to \texttt{Column.Feed[1]} (or appropriate feed stage)}\\
    \textit{\textbf{2. Connect Column to Products:}\\
    - Connect \texttt{Column.Condenser.Out} to \texttt{Distillate.In}\\
    - Connect \texttt{Column.Reboiler.Out} to \texttt{Bottoms.In}}\\
    \textit{After you make these connections, I can help you:\\
    - Set the feed stage location\\
    - Configure the column specifications (reflux ratio, distillate rate, etc.)\\
    - Run the simulation\\
    - Analyze the results}\\
    \textit{Let me know when you've made the connections and I'll continue with
    the configuration!}
\end{framed}

\begin{framed}
    \textbf{Prompt 2.1.1:}

    \textit{I did the connection, please set other configurations, but before
    you analyze the results, pause and let me set the setup to "Solve".}
\end{framed}

\begin{framed}
    \textbf{Answer 2.1.1:}

    \textit{Perfect! Now that you've made the connections, let me configure the
    remaining column specifications. First, let me check the current parameters
    to see what feed stage options are available:}\\
    $\longrightarrow$ \textit{\textbf{model\_all\_params}(model\_name: Column)\textrightarrow(success: true, ...)}\\
    \textit{Good! I can see the feed stage parameter \texttt{FeedStage[S1]}. Let
    me set the feed stage to approximately the middle of the column (stage 8) and
    check what variables are available for column specifications:}\\
    $\longrightarrow$ \textit{\textbf{param\_set\_multiple}(parameter\_data: [{path: Column.FeedStage[S1], value: 8}])\textrightarrow(success: true, ...)}\\
    \textit{Now let me check the column variables to see what specifications I
    can set:}\\
    $\longrightarrow$ \textit{\textbf{model\_all\_vars}(model\_name: Column)\textrightarrow(success: true, ...)}\\
    \textit{Perfect! I can see the key column variables. Now let me set typical
    column specifications for a water-methanol separation. I'll set:\\
    1. \textbf{Reflux Ratio} = 2.0 (reasonable for this separation)\\
    2. \textbf{Distillate Rate} = around 50\% of feed (since we want to separate
    into two products)\\
    3. \textbf{Contact Efficiency} = 1.0 for rigorous calculation}\\
    \textit{Let me calculate the distillate flow rate. With 10 kg/s feed and approximately
    equal molecular weights, I'll target about 50\% mass recovery in distillate:}\\
    $\longrightarrow$ \textit{\textbf{var\_set\_multiple}(variable\_data: [{path: Column.RefluxRatio, value: 2},
    {path: Column.Lmdist, unit: kg/s, \newline \hspace*{0.6cm} value: 5}, {path: Column.Contact, value: 1}])\textrightarrow(success: true, ...)}\\
    \textit{Now let me set the column pressure profile. I'll set a small pressure
    drop across the column:}\\
    $\longrightarrow$ \textit{\textbf{var\_set\_multiple}(variable\_data: [{path: Column.Ptop, unit: kPa, value: 101.325},
    {path: Column.DPcol, unit: \newline \hspace*{0.6cm} kPa, value: 2}])\textrightarrow(success: true, ...)}\\
    \textit{Let me check if the column has proper condenser and reboiler duties
    by getting key convergence variables:}\\
    $\longrightarrow$ \textit{\textbf{var\_get\_multiple}(variables: [Column.DutyCond,
    Column.DutyReb, Column.Tcond, Column.Treb,
    \newline
    \hspace*{0.6cm} Column.RefluxRatio, Column.Lmdist])\textrightarrow(success: true, ...)}\\
    \textit{Great! The column is showing reasonable values:\\
    - \textbf{Condenser Duty}: -18.8 MW (cooling required)\\
    - \textbf{Reboiler Duty}: +8.7 MW (heating required)\\
    - \textbf{Condenser Temperature}: \textasciitilde69.5°C\\
    - \textbf{Reboiler Temperature}: \textasciitilde76.1°C\\
    - \textbf{Reflux Ratio}: 2.0 as set\\
    - \textbf{Distillate Rate}: \textasciitilde4.6 kg/s}\\
    \textit{Now let me save the simulation:}\\
    \textit{$\longrightarrow$ \textbf{sim\_save}()\textrightarrow(success: true, ...)} \\
    \textit{Perfect! I have successfully configured the water-methanol distillation
    simulation with:}

    \textit{\textbf{Simulation Configuration Summary:}}\\
    \textit{\textbf{Feed Conditions:}\\
    - \textbf{Feed Rate}: 10 kg/s\\
    - \textbf{Temperature}: 80°C (353.15 K)\\
    - \textbf{Pressure}: 1 bar (101.325 kPa)\\
    - \textbf{Composition}: 50 mol\% Water, 50 mol\% Methanol}\\
    \textit{\textbf{Column Configuration:}\\
    - \textbf{Stages}: 15 theoretical stages\\
    - \textbf{Feed Stage}: 8 (middle of column)\\
    - \textbf{Condenser}: Internal condenser\\
    - \textbf{Reboiler}: Internal reboiler\\
    - \textbf{Thermodynamics}: NRTL method}\\
    \textit{\textbf{Operating Specifications:}\\
    - \textbf{Reflux Ratio}: 2.0\\
    - \textbf{Distillate Rate}: \textasciitilde4.6 kg/s (automatic based on specifications)\\
    - \textbf{Column Pressure}: 101.325 kPa (top) with 2 kPa pressure drop\\
    - \textbf{Contact Efficiency}: 1.0 (rigorous calculation)}\\
    \textit{\textbf{Current Performance:}\\
    - \textbf{Condenser Temperature}: \textasciitilde69.5°C\\
    - \textbf{Reboiler Temperature}: \textasciitilde76.1°C\\
    - \textbf{Condenser Duty}: -18.8 MW (cooling)\\
    - \textbf{Reboiler Duty}: +8.7 MW (heating)}

    \textit{\textbf{Now you can proceed to set the Setup parameter to "Solve" in
    the AVEVA interface to run the full rigorous calculation.}}\\
    \textit{After you change the Setup to "Solve" and the simulation converges, let
    me know and I'll help analyze the results including:\\
    - Product compositions and purities\\
    - Energy requirements\\
    - Column performance\\
    - Separation efficiency}\\
    \textit{Ready when you are!}
\end{framed}

\begin{framed}
    \textbf{Prompt 2.1.2:}

    \textit{I change to "Solve", now the simulation should have a different result.}
\end{framed}

\begin{framed}
    \textbf{Answer 2.1.2:}

    \textit{Excellent! Now that you've changed the setup to "Solve", let me
    analyze the converged simulation results. The column should now be running rigorous
    calculations instead of the initial estimates.}\\
    \textit{$\longrightarrow$ \textbf{sim\_status}()\textrightarrow(success: true, ...)}\\
    \textit{Great! The simulation shows as solved. Let me get the detailed
    column performance after the rigorous solve:}\\
    $\longrightarrow$ \textit{\textbf{var\_get\_multiple}(variables: [Column.DutyCond,
    Column.DutyReb, Column.Tcond, Column.Treb,
    \newline
    \hspace*{0.6cm} Column.RefluxRatio, Column.Lmdist, Column.Lmbtms, Column.Lreflux,
    Column.Fboilup])\textrightarrow(success: true, \newline
    \hspace*{0.6cm}...)}\\
    \textit{Now let me check the product compositions and temperatures:}\\
    \textit{$\longrightarrow$ \textbf{connectors\_list}()\textrightarrow(success: true, ...)}\\
    \textit{Perfect! Now let me get the composition and properties of the product
    streams:}\\
    $\longrightarrow$ \textit{\textbf{var\_get\_multiple}(variables: [S2.T, S2.P,
    S2.W, S2.z[Water], S2.z[Methanol], S3.T, S3.P, S3.W, S3.z[Water],
    \newline
    \hspace*{0.6cm} S3.z[Methanol], S1.T, S1.W, S1.z[Water], S1.z[Methanol]])\textrightarrow(success: true, ...)}\\
    \textit{Let me also get the stage-by-stage composition profile to see the
    separation efficiency:}\\
    $\longrightarrow$ \textit{\textbf{var\_get\_multiple}(variables: [Column.T[1],
    Column.T[8], Column.T[15], Column.VF[1], Column.VF[8],
    \newline
    \hspace*{0.6cm} Column.VF[15]])\textrightarrow(success: true, ...)}

    \textit{\textbf{RIGOROUS SIMULATION RESULTS ANALYSIS}}\\
    \textit{\textbf{Overall Material Balance:}\\
    - \textbf{Feed}: 10.0 kg/s (50\% Water, 50\% Methanol at 80°C)\\
    - \textbf{Distillate (S2)}: 5.68 kg/s (Methanol-rich product)\\
    - \textbf{Bottoms (S3)}: 4.32 kg/s (Water-rich product)\\
    - \textbf{Mass Balance}: 5.68 + 4.32 = 10.0 kg/s \checkmark}\\
    \textit{\textbf{Product Compositions:}}\\
    \textit{\textbf{Distillate (Overhead Product):}\\
    - \textbf{Temperature}: 64.3°C (337.5 K)\\
    - \textbf{Pressure}: 100.3 kPa\\
    - \textbf{Flow Rate}: 5.68 kg/s\\
    - \textbf{Composition}:\\
    \hspace*{.5cm}- \textbf{Methanol}: 99.49 mol\% (High purity!)\\
    \hspace*{.5cm}- \textbf{Water}: 0.51 mol\%}\\
    \textit{\textbf{Bottoms (Bottom Product):}\\
    - \textbf{Temperature}: 87.8°C (361.0 K)\\
    - \textbf{Pressure}: 102.3 kPa\\
    - \textbf{Flow Rate}: 4.32 kg/s\\
    - \textbf{Composition}:\\
    \hspace*{.5cm}- \textbf{Water}: 89.68 mol\%\\
    \hspace*{.5cm}- \textbf{Methanol}: 10.32 mol\%}\\
    \textit{\textbf{Energy Requirements:}\\
    - \textbf{Condenser Duty}: -18.85 MW (cooling required)\\
    - \textbf{Reboiler Duty}: +8.77 MW (heating required)\\
    - \textbf{Net Energy}: 8.77 MW heating input}\\
    \textit{\textbf{Column Temperature Profile:}\\
    - \textbf{Top Stage (1)}: 64.7°C - lightest components\\
    - \textbf{Feed Stage (8)}: 74.3°C - feed entry point\\
    - \textbf{Bottom Stage (15)}: 79.2°C - heaviest components}\\
    \textit{\textbf{Separation Performance:}}\\
    \textit{\textbf{Methanol Recovery:}\\
    - Feed methanol: \textasciitilde50 mol\% × 10 kg/s = significant methanol input\\
    - Distillate purity: \textbf{99.49\% methanol} - excellent separation!}\\
    \textit{\textbf{Water Recovery:}\\
    - Bottoms purity: \textbf{89.68\% water} - good but some methanol carryover}

    \textit{\textbf{Key Performance Indicators:}\\
    1. \textbf{Excellent methanol purity} in distillate (99.49\%)\\
    2. \textbf{Good water recovery} in bottoms (89.68\%)\\
    3. \textbf{Reasonable energy consumption} (8.77 MW reboiler duty)\\
    4. \textbf{Proper temperature gradient} across column stages\\
    5. \textbf{Stable operating conditions} with internal reflux/boilup}\\
    \textit{The simulation shows excellent separation performance with high-purity
    methanol overhead product and good water recovery in the bottoms. The energy
    requirements are reasonable for this type of separation.}
\end{framed}
\end{document}